\documentclass{aa}  

\usepackage{graphicx}
\usepackage{txfonts}
\usepackage{longtable}
\usepackage{multirow}
\usepackage{pifont}
\usepackage{comment}
\usepackage[dvipsnames]{xcolor}
\usepackage{array}
\usepackage{tabularx}
\usepackage{booktabs}
\usepackage{makecell}

\newcommand{\msun}{\mbox{M$_{\odot}$}}

\newcommand{\lsun}{\mbox{L$_{\odot}$}}
\newcommand{\teff}{\mbox{$T_\mathrm{eff}$}}
\newcommand{\Mz}{\mbox{$M_\mathrm{ZAMS}$}}
\newcommand{\MHe}{\mbox{$M_\mathrm{He}$}}
\newcommand{\MCO}{\mbox{$M_\mathrm{CO}$}}
\newcommand{\parsec}{\textsc{parsec}}
\newcommand{\Mh}{\mbox{\rm [{\rm M}/{\rm H}]}}
\newcolumntype{H}{>{\setbox0=\hbox\bgroup}c<{\egroup}@{}}

\usepackage[colorlinks]{hyperref}
\begin{document}

   \title{Evolutionary tracks, ejecta, and ionizing photons from intermediate-mass to very massive stars with PARSEC\thanks{Stellar tracks, ejecta, and ionizing photon tables are available in the \parsec\ database at \url{http://stev.oapd.inaf.it/PARSEC/}.}
   }

   \author{G. Costa\inst{1,2,3}\fnmsep\thanks{corresponding author: guglielmo.costa.astro@gmail.com}
          \and
          K. G. Shepherd\inst{4, 3}\fnmsep\thanks{corresponding author: kshepher@sissa.it}
          \and
          A. Bressan\inst{4}
          \and
          F. Addari\inst{4}
          \and
          Y. Chen\inst{5}
          \and 
          X. Fu\inst{6}
          \and
          G. Volpato\inst{1, 3}
          \and    
          C. T. Nguyen\inst{7,4}
          \and \\
          L. Girardi\inst{3}
          \and
          P. Marigo\inst{1}
          \and
          A. Mazzi\inst{8,1}
          \and
          G. Pastorelli\inst{1,3}
          \and
          M. Trabucchi\inst{1,3}
          \and
          D. Bossini\inst{1,3}
          \and
          S. Zaggia\inst{3}
          }

   \institute{
        Dipartimento di Fisica e Astronomia, Universit\`a degli studi di Padova,
        Vicolo dell'Osservatorio 3, Padova, Italy
    \and
        Univ Lyon, Univ Lyon1, ENS de Lyon, CNRS, Centre de Recherche Astrophysique de Lyon UMR5574, F-69230 Saint-Genis-Laval, France
    \and
        INAF - Osservatorio Astronomico di Padova, Vicolo dell'Osservatorio 5, Padova, Italy
    \and
        SISSA, via Bonomea 365, I--34136 Trieste, Italy
    \and
        Anhui University, Hefei 230601, China    
    \and
        Purple Mountain Observatory, Chinese Academy of Sciences, Nanjing 210023, China
    \and
        INAF - Osservatorio Astronomico di Trieste, Via Giambattista Tiepolo 11, Trieste, Italy
    \and
        Dipartimento di Fisica e Astronomia Augusto Righi, Universit\`a degli Studi di Bologna,
        Via Gobetti 93/2, 40129 Bologna, Italy
             }

   \date{Received 11 October, 2024; accepted 22 December, 2024}

\hypersetup{
    %draft,
    %colorlinks=true,
    linkcolor=blue,
    citecolor=blue,
    filecolor=magenta,      
    urlcolor=blue
}
 
\abstract
{
    Recent advancements in stellar evolution modeling offer unprecedented accuracy in predicting the evolution and deaths of stars.
    We present new stellar evolutionary models computed with the updated \parsec\ V2.0 code for a comprehensive and homogeneous grid of metallicities and initial masses.
    Nuclear reaction networks, mass loss prescriptions, and the treatment of elemental mixing have all been updated in \parsec\ V2.0. 
    We computed models for thirteen initial metallicities spanning $Z = 10^{-11}$ to $Z = 0.03$, with masses ranging from 2.0~\msun\ to 2000~\msun, consisting of a library of over 1,100 ($\sim 2100$ tracks including pure-He models) full stellar evolution tracks.
    For each track, the evolution is followed from the pre-main-sequence to the most advanced early-asymptotic-giant-branch or the pre-supernova phases (depending on the stellar mass).
    Here, we describe the properties of the tracks and their chemical and structural evolution. 
    We computed the final fates and the remnant masses and built the mass spectrum for each metallicity, finding that the combined black hole (BH) pair-instability mass gap spans just between 100 and 130 \msun.
    Moreover, the remnant masses provide models consistent with observed BH masses, such as those from the primaries of GW190521, Cygnus X-1, and \textit{Gaia} BH3 binary systems. 
    We computed and provided the chemical ejecta from stellar winds and explosive final fates, along with the ionizing photon rates.
    We show how metallicity affects the evolution, fates, ejecta, and ionizing photon counts from these stars. 
    Our results show strong overall consistency with other tracks computed with different codes, and the most significant discrepancies arise for very massive stars ($\Mz\ > 120$~\msun) due to the different treatment of mixing and mass loss.
    A comparison with a large sample of observed massive stars in the Tarantula Nebula of the Large Magellanic Cloud shows that our tracks nicely reproduce the majority of stars that lie on the main sequence.
    All the models are publicly available and can be retrieved in the \textsc{parsec} database.
  }

   \maketitle
%
%-------------------------------------------------------------------
\section{Introduction}
\label{sec:Intro}

Stellar evolutionary models are essential building blocks needed to study many astrophysical problems, such as determining the cosmological Hubble constant with Cepheids \citep{Riess2019}, deriving the age of the Universe with globular clusters \citep{Valcin2020}, understanding how galaxies form and evolve \citep{Rosdahl2018} and their spectral evolution \citep{Bruzual2003, Lecroq2024}, investigating the Universe's chemical enrichment \citep{Kobayashi2020}, understanding the final fates of star's \citep{Woosley2017, Limongi2018, Marigo2020}, study the formation scenarios of the binary black holes (BHs) discovered via gravitational waves \citep{Zevin2021, Iorio2023}, and determining the exoplanets' host stars' properties \citep{Tayar2022}.

Stellar models play an important role in the interpretation and analysis of a large (and increasing) amount of astrophysical data. Therefore, there are several large grids of tracks available in the literature that have been computed with various stellar evolutionary codes, like \textsc{yrec} \citep{Spada2017}, \textsc{mist} \citep[][with \textsc{mesa}]{Choi2016}, \textsc{franec} \citep{Limongi2018}, \textsc{basti} \citep{Hidalgo2018}, \textsc{starevol} \citep{Amard2019}, \textsc{genec} \citep{Ekstrom2012, Georgy2013, Murphy2021, Sibony2024}, the \textsc{Bonn} code \citep{Brott2011, Szecsi2022}, and \textsc{parsec} \citep{Bressan2012, Chen2014, Tang2014, Chen2015, Nguyen2022}, to name a few.
However, they all have different assumptions for the input physics, initial mass range, and metallicity span.

It is paramount to continuously update and release state-of-the-art homogeneous sets of stellar tracks and make them available to the research community.
In this paper, we adopt the new version of the PAdova and tRieste Stellar Evolution Code \citep[\textsc{parsec v2.0},][]{Costa2019a, Nguyen2022} to compute new large grids of stellar evolutionary tracks ranging from intermediate mass to massive and very massive, with a metallicity ($Z$, in mass fraction) that spans from primordial chemical abundances to super-solar values. 
Since we present several sets of tracks with a large range in metallicity, it is convenient to group stars into three large families: metal-free, metal-poor, and solar-like populations.

The first group comprises the first stars to have formed in the Universe \citep[at a redshift $\lesssim 30$,][]{Klessen2023}. 
These are often called Population III (or Pop III) stars.
They were formed from the collapse of the almost metal-free primordial gas clouds, with just a tiny amount of lithium left by the Big Bang nucleosynthesis.
They contributed to the beginning of the Universe's reionization process and early chemical enrichment, mostly via supernovae ejection \citep[but, possibly, also via pulsation-driven mass loss, e.g.,][]{Volpato2023, Volpato2024}.
Due to the lack of direct observations, the physical details and properties of the formation and evolution of these stars are a topic of ongoing discussion. 
Although it is not possible to compare models directly with observed Pop III stars, these models are advantageous for investigating the integrated properties of this population; for example, ionizing fluxes and chemical ejecta. 
It is commonly accepted that the scarcity of metals favors Pop III stars having a more top-heavy stellar initial mass function (IMF) than their metal-richer counterparts \citep[e.g.,][]{Hirano2014, Stacy2016, Chon2021, Park2023}, and that Pop III stars do not lose mass via stellar winds during their evolution. Therefore, they are believed to be possible progenitors of massive stellar-mass BHs
\citep[e.g.,][]{Belczynski2017, Kinugawa2020, Tanikawa2022, Costa2023, Santoliquido2023}, in agreement with the large BH masses detected via gravitational waves by the LIGO--Virgo--KAGRA collaboration interferometers \citep[][]{abbottGWTC2.1,abbottGWTC3}, and also to be possible seeds of intermediate-mass BHs ($\sim 10^2 - 10^5~\msun$).

The second generation of stars, also known as Population II (or Pop II), formed from the interstellar medium (ISM), enriched by the explosions of Pop III stars. 
Pop II stellar formation begins to dominate the cosmic star formation rate density around $z \sim$ 15 \citep{Klessen2023}, and since the metal-enriched gas has more possible cooling mechanisms, Pop II stars are thought to be able to form a higher number of lower-mass, longer-lived stars than Pop III. 
In fact, Pop II stars are the oldest directly observable stars in the Universe (so far), observed in the halo of the Milky Way, in metal-poor globular clusters, and dwarf galaxies \citep[e.g.,][]{Frebel2015, Frebel2019, Gratton2019}. 
At low metallicity, Pop II stars have weak stellar winds, and thus they can still retain most of their mass until the end of the evolution, eventually forming massive stellar BHs \citep{Costa2023}.

The last generation of stars, Population I (or Pop I) stars, were born from the enriched ashes of the intermediate and massive Pop II stars.
They are the major stellar component of the Milky Way and Magellanic Clouds.
The chemical composition of Pop I stars provides important clues to the local star formation history and nearby stellar evolution. 
Massive Pop I stars are characterized by strong stellar winds, which makes them important polluters of the ISM in the local universe. 
The IMF of Pop I stars is well constrained by several measurements \citep{Salpeter1955, Kroupa2001, Chabrier2003}, which find the great majority of stars to belong to the low-mass regime, yet there is still debate about the maximum initial mass that they can form.
Recent observations of stars in the Tarantula Nebula of the Large Magellanic Cloud (LMC) have found masses up to $\sim$ 200$-$300 \msun\ \citep{Crowther2010, Brands2022}, setting important constraints on the stellar winds prescriptions adopted by models.
These constraints are also important to derive the maximum mass of BHs at solar metallicity \citep[e.g.,][]{Bavera2023, Romagnolo2024, Vink2024}.

This paper is structured as follows. 
In Sec.~\ref{sec:Meth}, we describe the input physics assumed in our new \textsc{parsec} stellar tracks, describing with particular care the updates with respect to the previous release of tracks \citep[by][]{Chen2015}.
We present the new grids of tracks in Sec.~\ref{sec:Res}. 
In particular, we describe their path in the Hertzsprung-Russell (HR) diagram in Sec.~\ref{subsec:HR_diag}; the structure and chemical evolution of selected massive stars in Sec.~\ref{subsec:struct_evol}; the stellar final fates and mass spectrum in Sec.~\ref{subsec:Fates}; 
the derived stellar total ejecta in Sec.~\ref{subsec:Ejecta}; and finally, in Sec.~\ref{subsec:Ioniz_rad}, the ionizing photon emission of these stars. 
In the discussion, Sec.~\ref{sec:discussion}, we undertake a comparison with other codes and with observations, and in Sec.~\ref{sec:conclusion}, we derive our conclusions.
The appendices provide more details about the stellar models' properties and the grids of pure-helium (or pure-He) stellar tracks. 
All the tables of stellar tracks, plus the ejecta and ionizing photons, are publicly available in our \parsec\ web database\footnote{\url{http://stev.oapd.inaf.it/PARSEC/}}.

The sets of tracks published here have already been successfully applied to several studies \citep[such as][]{DalTio2022, Costa2023, Santoliquido2023, Lecroq2024, Liu2024, Mazzi2024, Mestichelli2024}, and represent the current standard evolutionary library for the \textsc{sevn} \citep{Iorio2023} and \textsc{trilegal} \citep{DalTio2021} population synthesis codes.

%--------------------------------------------------------------------
\section{Input physics and code description}
\label{sec:Meth}
  
In this work, we present new grids of non-rotating \parsec\ stellar tracks with a range of initial mass from 2 to 600 \msun\ for $Z > 10^{-4}$ and up to 2000 \msun\ for $Z \leq 10^{-4}$, from the pre-main-sequence (PMS, described in App.~\ref{app:PMS}) to the most advanced burning phases, as is described below.
  
Intermediate-mass stars ($2 \leq \Mz/\msun \leq 9$) are evolved through all the main burning phases up to the early asymptotic giant branch (AGB) phase.
The computation of the most advanced thermally pulsating phases requires careful treatment and goes beyond the scope of this paper.
Nonetheless, in a recent work, we updated \parsec\ in all the input physics and numerical treatments needed to follow the complete evolution of low- and intermediate-mass stars at solar metallicity, including the thermally pulsating AGB phase \citep[see][]{Addari2024}.
The computation of an extensive grid of full AGB stars for different metallicities is deferred to a forthcoming study. 

Massive stars ($\Mz \gtrsim 10$) are followed until the advanced pre-supernova oxygen burning phase or until the occurrence of the electron-positron (e$^+$ - e$^-$) pair instability (PI) regime.

This new version of \parsec\ includes several updates with respect to the latest release of massive stars by \citet{Chen2015} using \textsc{parsec v1.2s}.
These updates have been incorporated in the code step-by-step in several works \citep[i.e.,][]{Fu2018, Costa2019a, Costa2019b, Costa2021, Nguyen2022}. 
In the following, we briefly describe the physics adopted in the new stellar models.

\subsection{Opacities, equation of state, nuclear reactions and overshooting}
\label{subsec:micro_phys}

Opacities in the high-temperature regime ($4.2 \leq \log (\teff/\mathrm{K}) \leq 8.7$) are taken from \textsc{OPAL}\footnote{\url{https://opalopacity.llnl.gov}} \citep[Opacity Project At Livermore,][]{Iglesias1996}, while for the low-temperature regime ($3.2 \leq \log (\teff/\mathrm{K}) \leq 4.1$) we adopt the \textsc{\ae sopus}\footnote{\url{http://stev.oapd.inaf.it/aesopus}} tool from \citet{Marigo2009}. 
Between the high and low-temperature regimes, opacities are interpolated between the \textsc{\ae sopus} and \textsc{OPAL} values.
Conductive opacities are included from \citet{Itoh2008}. 

The equation of state is calculated from the \textsc{freeeos} code developed by A. W. Irwin\footnote{\url{http://freeeos.sourceforge.net/}}.
The \textsc{freeeos} code does not account for pair creation, so for temperatures $\log (T/\mathrm{K}) > 8.5$, we adopt the code by \citet{Timmes1999}.

The nuclear reaction network solves for the abundances of 32 isotopic elements from hydrogen to zinc (plus neutrons) and considers 72 reactions in total. 
The network consists of all important reactions from hydrogen to oxygen burning; that is, p-p chains, CNO tricycle, Ne-Na, and Mg-Al chains, and relevant $\alpha$-capture reactions (and their inverse) up to $^{56}$Ni (a full list of the reaction network can be found in \citealp{Fu2018} and \citealp{Costa2021}).
Reaction rates and Q-values are taken from the \textsc{jina reaclib} database \citep{Cyburt2010}. 
In \textsc{parsec} v2.0, the nuclear reaction network and element mixing are solved at the same time by adopting an implicit diffusive scheme \citep[][]{Marigo2013, Costa2019a}. 

Energy losses from electron neutrinos are taken from \citet{Munakata1985} and \citet{ItohKohyama1983}. A fitting formula from \citet{Haft1994} is used for plasma neutrinos.

Convective energy transport is described using the mixing-length theory \citep{Bohm-Vitense1958}, and we set the value of the mixing-length parameter $\alpha_\mathrm{MLT} = 1.74$ \citep[as was suggested by the solar calibration by][]{Bressan2012}. 
Stability against convection of radiative zones is tested using the Schwarzchild criterion \citep{Schwarzschild1958}.
We include core overshooting, allowing for the penetration of convective elements into stable regions \citep[penetrative overshooting, described by][]{Bressan1981}. 
The overshooting parameter is $\lambda_{\mathrm{ov}} = 0.5$ (in pressure scale heights, $H_P$) and corresponds to the maximum distance traveled by the overshooting bubble across the unstable border. 
Since $\lambda_{\mathrm{ov}}$ is taken to be roughly halfway across the border, the core overshooting length (or overshooting distance) is $l_{\mathrm{ov}} \approx 0.5~\lambda_{\mathrm{ov}}~H_P = 0.25~H_P$.
This value roughly corresponds to $f_\mathrm{ov} = 0.025$ in the exponential decay overshooting formalism \citep{Herwig1997}, which is often used by other groups \citep[e.g.,][]{Choi2016}.
We also take into account the overshooting from the convective envelope, which extends from the base of the convective envelope downward. 
In this case, we assume a step overshooting region that extends by $\Lambda_{\mathrm{env}}$ = 0.7 $H_P$ from the base of the convective envelope. 

Another important thing to consider when modeling massive stars is the density inversion that could occur in the low-density convectively inefficient envelope of red supergiants (RSGs) when the local luminosity of the layer becomes higher than the Eddington luminosity.
A density inversion can result in numerical instabilities that make it difficult to follow the evolution of such stars.
Almost all stellar evolutionary codes adopt ad hoc strategies to cure density inversions \citep[as is nicely overviewed by][]{Agrawal2022}.
In \textsc{parsec v2.0}, we adopt the same strategy for massive stars as is described in \citet{Chen2015}, and already proposed by \citet{Bressan1993}.
The treatment consists of imposing a limit to the real temperature gradient $\nabla_{T} \leq \nabla_{Tmax} = \frac{1 - \chi_{\mu} \nabla_{\mu}}{\chi_{T}}$, where $\chi_{\mu}$ and $\chi_{T}$ are thermodynamic derivatives and $\nabla_{\mu}$ is the molecular weight gradient.
This formalism takes to a more efficient convection that helps prevent the rise of a density inversion \citep[and, in general, inhibits the envelope inflation,][]{Sanyal2015}. 
While necessary to prevent numerical difficulties, it should be kept in mind that this approach has been known to result in slightly smaller radii and hotter \teff\ compared with RSG models that allow for a density inversion \citep[][]{Chen2015, Agrawal2022}.

\begin{table*}
    \caption{Initial composition in mass fraction and mass range of the new sets of evolutionary tracks.}
    \begin{center}
    \resizebox{\textwidth}{!}{\begin{tabular}{c|cccHcccccccccc}
        \toprule
        Population & \multicolumn{1}{c|}{\makecell{III}} & \multicolumn{6}{c|}{II} & \multicolumn{6}{c}{I} \\
        \midrule
        $X$ & 0.751 & 0.751 & 0.751 & 0.750 & 0.749 & 0.746 & 0.740 & 0.735 & 0.729 & 0.724 & 0.713 & 0.704 & 0.696 & 0.668\\
        $Y$ & 0.2485 & 0.2485 & 0.2487 & 0.249 & 0.250 & 0.252 & 0.256 & 0.259 & 0.263 & 0.267 & 0.273 & 0.279 & 0.284 & 0.302\\
        $Z$ & $10^{-11}$ &  $10^{-6}$ &  $10^{-4}$ &  $5 \times 10^{-4}$ &  0.001 &  0.002 &  0.004 &  0.006 &  0.008 & 0.01 &  0.014 &  0.017 &  0.02 & 0.03 \\
        \Mh$^\dagger$ & -9.192 & -6.192 & -4.192 & -1.492 & -1.190 & -0.888 & -0.583 & -0.404 & -0.276 & -0.176 & -0.023 & +0.067 & +0.143 & +0.336 \\
        %\hline
        \midrule
        \makecell{\Mz\ range \\ $[\msun]$} & \multicolumn{3}{c|}{[2 - 2000]} & \multicolumn{11}{c}{[2 - 600]} \\
        % \hline
        \bottomrule
    \end{tabular}}
    \end{center}
    \footnotesize{$^\dagger$ Approximated value from $\Mh = \log ((Z/X)/0.0207)$ \citep{Bressan2012}.}
    \label{tab:initial_comp}
\end{table*}

% ---------------------------- mass loss -----------------------------
\subsection{Mass loss}
\label{subsec:Mloss}

In our computations, we include mass loss via stellar winds prescriptions depending on the mass and evolutionary phase. Details on the winds' implementation in \parsec\ can be found in \citet{Chen2015}, \citet{Costa2021} and \citet{Nguyen2022}. 
In the following, we provide a short summary of the adopted winds prescriptions.

For intermediate-mass stars, we include stellar winds by using the prescription by \citet{Reimers1975}, with an $\eta_R = 0.2$ as is described in \citet{Nguyen2022} and derived from asteroseismic measurements of stars in Galactic open clusters \citep{Miglio2012}. 
This mass loss is used in the red giant branch (RGB) and the core-helium burning (CHeB) phase. 
For the mass loss in post-CHeB phases, we adopt the new wind recipe described by \citet{Marigo2020} and \citet{Pastorelli2020}, in which the AGB wind is treated as a two-stage process \citep[included in \textsc{parsec} by][]{Addari2024}. 
The first stage is the dust-free mass loss regime for the E-AGB phase, in which we adopt the prescription by \citet{Cranmer2011}\footnote{This formulation takes into account the effect of magnetic fields and rotation on the cool stellar atmosphere.}. 
In the second stage, higher luminosities and lower effective temperatures favor the generation of strong stellar winds driven by the interaction of radiation with condensed dust grains. 
For this dust-driven mass loss regime, we adopt different prescriptions depending on the surface carbon-to-oxygen (C/O) ratio. 
For AGB stars with C/O$ < 1$, we adopt the formula by \citet{Bloecker1995}. 
On the other hand, for carbon-rich AGB stars with C/O$ > 1$, still unable to form carbon dust efficiently, we use mass loss by \citet{Winters2000}. 
Finally, for AGB stars able to form significant amounts of carbonaceous dust, we adopt the prescriptions by \citet{Mattsson2010}, \citet{Eriksson2014}, and \citet{Bladh2019}, obtained from dynamical atmosphere models.

For massive hot stars (\teff\ $\geq 10,000$ K), we use the mass loss prescriptions by \citet{Vink2000, Vink2001}, which include a metallicity dependence driven by the surface iron abundance, which reads $\dot{M} \propto (Z/Z_\odot)^{0.85}$, where $Z$ is the initial metallicity and $Z_\odot$ is the solar one. 
We also include the dependence on the Eddington ratio \citep{Grafener2008, Vink2011}, which becomes important for the most luminous stars. 
For cold massive stars (i.e., RSGs), we use the prescription by \citet{deJager1988}, including also the same metallicity adopted for the \citet{Vink2001} one.

For Wolf-Rayet (WR) stars (i.e., when the hydrogen surface abundance is $X\mathrm{^S_H} < 0.3$ in mass fraction, and $\log \teff \geq 4$), 
we use a luminosity-dependent prescription for the mass loss from \citet{Sander2019}, 
\begin{equation}
\log \frac{\dot{M}}{\msun \mathrm{yr^{-1}}} = -8.31 + 0.68 \cdot \log \frac{L}{\lsun} ~ ,
\label{eq:mloss_sander}
\end{equation}
which well represents galactic WR type-C (WC) and WR type-O (WO) stars. 
Since this empirical relation omits any metallicity dependence,  we use results from \citet{Vink2015} to account for the effects of metals in mass loss rates of WR type stars \citep[see][for more details]{Costa2021}. 

\subsection{Pair instability stopping condition}
\label{subsec:pairs}

During the advanced phases of stellar evolution, typically during the O burning, but it can also occur during C and Ne burning, the central temperatures and densities may be high enough that electron-positron pairs begin to be produced efficiently. 
The pair creation absorbs part of the plasma's total pressure, transformed in the rest mass of the created particles, thus inducing a dynamical instability called PI \citep{WoosleyHeger2002}.
As the central pressure is lowered, the star becomes dynamically unstable, and the core collapses. 
Afterward, oxygen is explosively ignited, leading to shock waves that propagate from the core to the envelope. 
These waves can, in turn, lead to external layer ejections or, in the extreme case, to the star's destruction \citep{Woosley2017}. 
The outcome of this instability depends heavily on the structure configuration, in particular on the mass of the carbon-oxygen core and its composition, as a greater amount of nuclear fuel (i.e., oxygen) destabilizes the star and drives more powerful explosions to reverse the collapse \citep{Farmer2019, Farmer2020, Woosley2021}.

As a hydrostatic stellar evolutionary code, \parsec\ cannot follow the dynamical phase induced by the PI. 
Therefore, to check when evolved massive stars become dynamically unstable, we use the \citet{Stothers1999} criterion, which gives a good approximation to determine the stellar stability without performing a detailed hydrodynamical simulation.
A star is considered stable if
\begin{equation}
\label{eq:PI}
     \langle \Gamma_1 \rangle = \frac{\int_0^{M_\mathrm{*}} \frac{\Gamma_1 P}{\rho}{}{dm}}{\int_0^{M_\mathrm{*}} \frac{P}{\rho}dm} > \frac{4}{3},
\end{equation}
where $\Gamma_1$ is the first adiabatic exponent, $P$ is the pressure, $\rho$ is the gas density, and $dm$ is the star element of mass.
The integral is computed from the center of the star ($m = 0$) up to the surface ($m = M_\mathrm{*}$, i.e., accounting for the current total mass of the star).
At each time step, we computed $\langle \Gamma_1 \rangle$ and stopped the evolution if it went below 4/3. 
A posteriori, when analyzing the track, we considered the star to be dynamically unstable if $\langle \Gamma_1 \rangle$ < 4/3 + 0.01, where 0.01 is added to be conservative \citep[as is suggested from dynamical investigations by][]{Marchant2019}. 
This approach has already been adopted to identify when a star enters the PI regime in previous works \citep{Costa2021, Costa2022, Volpato2023, Volpato2024}.

%-----------------------------------------------------------------

\section{Results}
\label{sec:Res}

These new sets of tracks extend and supersede the previous release computed with \parsec\ \textsc{v1.2s} \citep{Chen2015}, not only because of the most updated physics but also for the wider mass and metallicity ranges. 
The stellar tracks were computed for thirteen different initial metallicities, from $Z = 10^{-11}$ to $Z = 0.03$, as is listed in Table~\ref{tab:initial_comp}.
The initial helium content for each metallicity is determined by $Y = \frac{\Delta Y}{\Delta Z}Z + Y_\mathrm{p}$, where $Y_\mathrm{p}$ is the primordial helium abundance, taken to be 0.2485 from \citet{Komatsu2011}, and the helium to metal enrichment ratio $\Delta Y / \Delta Z = 1.78$ is based on solar calibration from \citet{Bressan2012}.
In this work we take the Solar metallicity to be $Z_{\odot} = 0.01524$, and adopt solar-scaled mixtures from \citet{Caffau2011}.
The investigation of alpha-enhanced mixtures for low metallicity populations will be a matter of future work. 

In this paper, we group our results based on metallicities that share typical features, and for clarity, we decide to use the nomenclature of Pop III, Pop II, and Pop I as follows. 
We identify our tracks with initial metallicity  $Z = 10^{-11}$ as Pop III stars. 
This is justified by the evolutionary features, characteristic of extremely metal-poor stars, which are found to emerge at $Z \leq 10^{-10}$ by several studies \citep[][]{Cassisi1993, Marigo2001, Ekstrom2008}.
It is worth noting that we computed our Pop III models with a primordial lithium of $\sim 2.7\times 10^{-9}$ in mass fraction, derived from the number fraction ratio over hydrogen  $5.2\times 10^{-10}$ given by \citet{Fields2011}.
Therefore, the total metal content (including Li) is $\sim 2.7\times 10^{-9}$, and the value $Z = 10^{-11}$ refers to the sum of all other metals than Li.
We refer to Pop II stars as those that form 
with initial metallicities from $Z = 10^{-6}$ to 0.004.
Finally, Pop I stars are assumed to have initial metallicities $Z \geq 0.006$.
In the following section, we describe the properties of each stellar population.

For each metallicity, we computed models from 2 to 600 ~\msun, while for $Z \leq 0.0001$, the models were extended up to 2000 \msun.
The mass ranges and mass step sizes are listed in table \ref{tab:mass_grid}. The grid has been spaced to produce a very fine grid of massive star models, particularly at masses where PI is expected to occur. 

\begin{table}%[h!]
    \caption{Grid of \Mz, and mass step. See text for more details.}
    \begin{center}
    \begin{tabular}{c|c}
        \toprule
        Initial Mass Range & Step Size  \\
         $[\msun]$ & $[\msun]$ \\
        \midrule
        2.0 - 10.0 & 0.2 \\
        10.0 - 12.0 & 1.0 \\
        12.0 - 30.0 & 2.0 \\
        30.0 - 80.0 & 10.0 \\
        85.0 - 100.0 & 5.0 \\
        100.0 - 300.0 & 10.0 \\
        300.0 - 800.0 & 50.0 \\
        800.0 - 1000.0 & 100.0 \\
        1000.0 - 2000.0 & 250.0 \\
        \bottomrule
    \end{tabular}
    \end{center}
    \label{tab:mass_grid}
\end{table}

\subsection{Stars on the Hertzsprung-Russell diagram}
\label{subsec:HR_diag}

\begin{figure*}[h!]
    \centering
    \includegraphics[width=\textwidth]{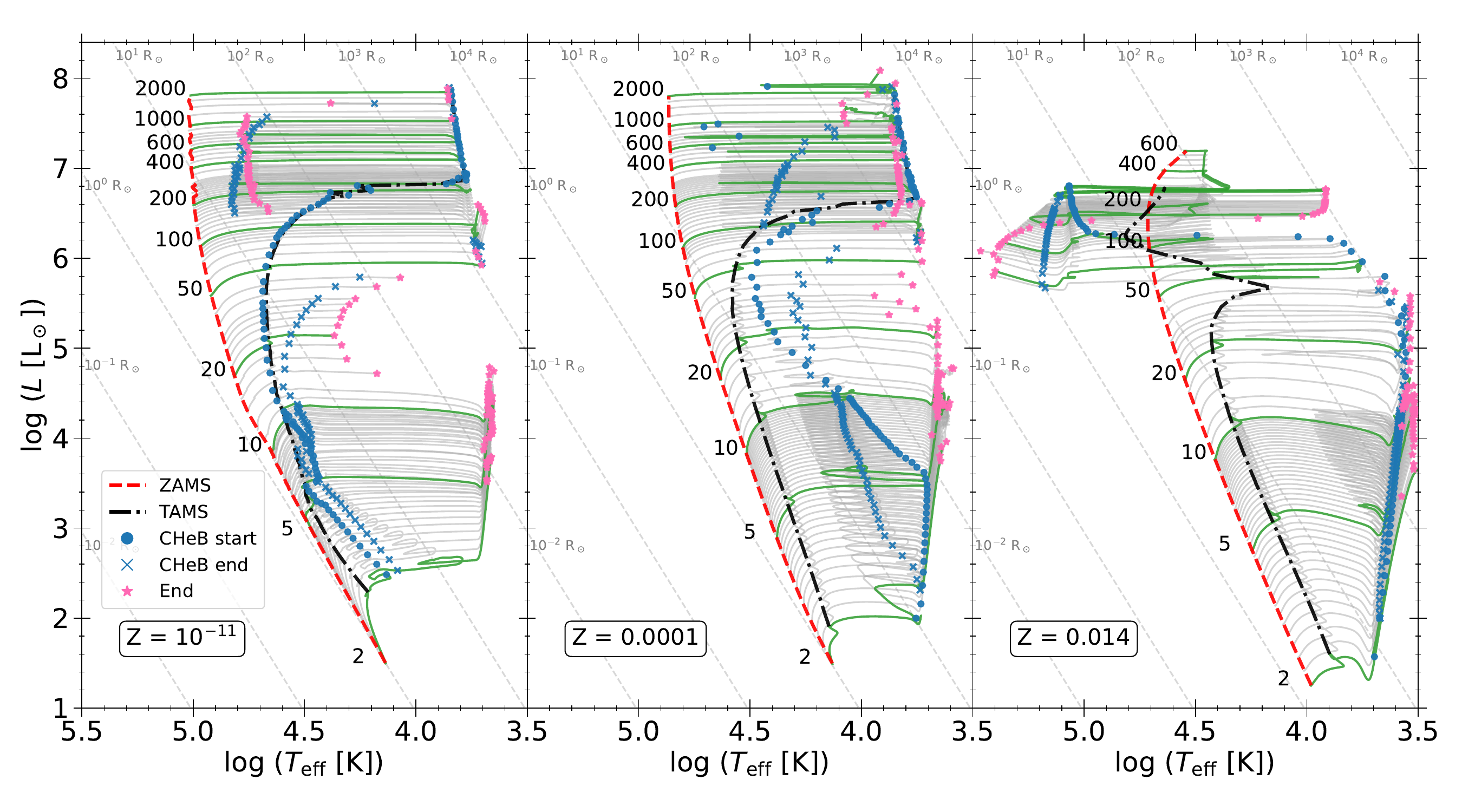}
    
    \caption{HR diagrams of tracks computed with $Z = 10^{-11}$, $Z = 0.001$ and $Z = 0.014$, in the left-hand, center, and right-hand panels, respectively. Tracks are shown beginning from the ZAMS, indicated by the dashed red line. The dash-dotted black line indicates the TAMS.
    Blue dots (crosses) indicate the beginning (end) of core helium-burning, respectively.
    Pink stars indicate the final position of the star.
    The pre-MS phase is not shown for clarity. 
    The stellar mass in solar masses of some selected tracks indicated by the solid green lines is written near their corresponding ZAMS position. Diagonal dashed gray lines show constant stellar radii as labeled.}
    \label{fig:HRD_Z}
\end{figure*}

\subsubsection{Pop III stars}
\label{subsec:popIII}

\begin{table}
  \centering
\caption{Definitions used for different stellar types.}
  \begin{tabular}{c|cc}
  \toprule
    \multicolumn{1}{c|}{Type}&\multicolumn{1}{c}{$\log \teff$} & \multicolumn{1}{c}{$X\mathrm{_H^S}$}\\ 
    \midrule
    WR & $\geq$ 4.0 & $<$ 0.3 \\ 
    BSG & $\geq$ 4.0 \\ 
    YSG & 3.7 - 4.0  \\
    RSG & $\leq$ 3.7 \\ 
    \bottomrule

  \end{tabular}

  \label{tab:type_defs}
\end{table}

The left-hand panel of Fig.~\ref{fig:HRD_Z} shows the Hertzsprung-Russell diagram of our Pop III tracks, starting from the zero-age main sequence (ZAMS) to the end of the computation, which depends on the initial mass, which is described in Sec.~\ref{sec:Meth}. 
Although we also computed the PMS phase of these tracks\footnote{In the web database, the evolutionary tracks include all their phases (also the PMS).}, here the PMS is omitted for the sake of clarity.
The PMS construction and some plots are described in App.~\ref{app:PMS}.
Pop III stars are more compact and hotter than their correspondent metal-rich stars.
This is mainly because of the lack of metals, as Pop III stars cannot ignite the CNO cycle during the early main sequence. 
Consequently, they contract until the central temperature (T$\mathrm{_c}$) is high enough that the p-p chain generates enough energy to sustain the star.
In massive Pop III stars, T$\mathrm{_c}$ could rise high enough to synthesize carbon from the 3-$\alpha$ reaction in the MS.
This may lead to the ignition of the CNO cycle, which replaces the p-p chain as the main energy source of the star \citep{Marigo2001, Murphy2021}. 

After the MS phase, Pop III stars have a very fast and smooth transition to the core He burning (CHeB) phase, mainly due to the already high central temperatures at the terminal age of the MS (TAMS).
Stars with $\Mz \leq 200~\msun$ finish their MS as yellow or blue supergiants (YSG or BSG), while those with $\Mz > 200~\msun$ end their MS as RSGs. The various stellar type definitions used in this work are shown in table \ref{tab:type_defs}.
In any case, stars ignite He at about the same location as the TAMS in the HR diagram (as is shown in Fig.~\ref{fig:HRD_Z}).
During the CHeB, stars with $\Mz \leq 100~\msun$ move to colder effective temperatures, while those with $\Mz > 100~\msun$ move back to the blue side of the diagram, becoming again BSGs.
At the end of CHeB, the core stops being convective and begins to contract, increasing the central temperature and density. 
When the internal conditions are met, stars ignite carbon.
The stars' position in the HR diagram does not change much during these advanced phases, and they die in about the same location, where they deplete carbon in their cores.
After central carbon depletion, neon photo-disintegration and then oxygen burning begin. 
Oxygen is ignited in the cores of models with $11 < \Mz/\msun < 100$, 
and the computation stops when almost all the central oxygen is depleted.
For $\Mz > 100~\msun$, tracks become unstable to PI (thus, the computation is stopped) shortly before core O-burning or even during the core C-burning, depending on the initial mass.
The final phases are further described in Sec.~\ref{subsec:Fates}.

The final position of stars in the HR diagram strongly depends on their initial mass.
Our tracks with a mass $\Mz \leq 14~\msun$ end their evolution in the red side of the diagram.
Those with $14 < \Mz/\msun < 50$ end as YSGs or BSGs \citep[other authors, e.g.,][have obtained similar results]{Marigo2001, Tanikawa2022}, while tracks in the interval $50 \leq \Mz/\msun \leq 100$ end their evolution as RSGs. 
Stars with an initial mass $100 \leq \Mz/\msun \leq 1000$ explode as BSGs.
Finally, stars with $\Mz > 1000~\msun$ end their life as RSGs.

\subsubsection{Pop II stars}
\label{subsec:PopII}

The center panel of Fig.~\ref{fig:HRD_Z} shows the HR diagram of a set with $Z = 0.0001$, representing Pop II stars.
The ZAMS of these stars is slightly cooler than Pop III, and they remain on the MS for a longer time (as is shown in Fig.~\ref{fig:app_lifetimes} and Table~\ref{tab:stellar_properties} in Appendix~\ref{app:Stellar_prop} and \ref{app:extra_tables} for a comparison).
At variance with metal-free stars, Pop II has enough metals to smoothly ignite the CNO cycle during the MS when the internal conditions are met.
Stars with $\Mz \lesssim 100~\msun$ end their MS with $\log \teff > 4$, while those with $\Mz \gtrsim 100~\msun$ tend to reach the TAMS as RSGs.
After the end of the MS, the stellar cores contract, and the envelopes expand, causing the stars to evolve redward.
Stars with masses $\Mz \lesssim 5~\msun$ reach the red giant branch before igniting helium, and during the CHeB, they perform the so-called blue loops.
Finally, they go back to the red and ascend the AGB.
As the initial mass increases, between 5 and 50~\msun, helium ignitions take place in bluer positions of the HR diagram, and the blue loops during the CHeB gradually become less extended until they disappear.
The position in the diagram of He ignition starts to go to colder \teff\ for stars with $\Mz > 50~\msun$.
As stars ignite helium in redder locations, blue loops during the CHeB re-appear.
Some very massive stars ($\Mz = 450$, 600, 750, 800, and 2000~$\msun$) move to the blue after the MS and ignite He as BSGs. 
After the CHeB, all tracks return to cold temperatures, and they evolve through all the advanced burning phases in the same location of the diagram until the end of their life.
At variance with Pop III stars, all tracks with $Z = 0.0001$ end their evolution on the cool side of the HR diagram. 

\subsubsection{Pop I stars}

As a representative case of the Pop I stars, in the right-hand panel of Fig.~\ref{fig:HRD_Z}, we show the HR diagram for $Z = 0.014$ stars, from the ZAMS to the advanced evolutionary phases. 
These stars start the MS being cooler and less compact than their corresponding lower metallicity stars. 
The evolution of massive stars shows the peculiar patterns of WR stars, which are not present in Pop III and Pop II stars due to their lack of strong stellar winds.
WR stars are luminous and hot and have atmospheres with a low hydrogen content ($X\mathrm{^S_H}$ < 0.3).

Stars with $2 \leq \Mz/\msun < 40~$ show a similar evolutionary path;
after they reach the TAMS, they move redward, crossing the Hertzsprung-Gap to the RGB, where they ignite helium and then perform the blue loops (although less extended than their lower metallicity counterparts).
After the CHeB phase, they enter into their final evolutionary stages (i.e., AGB for tracks below about 9 \msun\ and the oxygen-burning phase for the more massive stars) and move toward higher luminosity in the HR diagram.

The evolution of stars with $\Mz > 50~ \msun$ follows a different evolutionary pattern. 
In the early MS, the star follows the usual path of increasing luminosity and decreasing \teff. 
But during the MS, the star begins to lose a large amount of mass due to the strong stellar winds, which progressively peel off the star's outer layers and expose its hotter and He-rich inner layers.
When this happens, the star inverts its path, going toward higher \teff.
Depending on the initial mass, stars end their evolution in different places in the diagram.
Stars with $50 \leq \Mz/\msun \leq 190$ enter their pre-supernova phase as very hot BSGs ($\log \teff \geq 5$).
Stars with $\Mz > 190~\msun$, after the CHeB phase, evolve to lower \teff\, ending their lives as YSG stars.

\subsection{Structure and chemical evolution of very massive stars}
\label{subsec:struct_evol}

The evolution of very massive stars is profoundly sensitive to the initial metallicity. 
To highlight this, we consider the representative case of three tracks with \Mz\ = 240 \msun\ with distinct initial metallicities to examine how they differ in terms of structure and chemical composition throughout their evolution.
In particular, we show the role of stellar winds in shaping the evolution of stars, which in turn impact their final fates (discussed in Sec.~\ref{subsec:Fates}) and their chemical ejecta (described in Sec.~\ref{subsec:Ejecta}).

\begin{figure*}
    \centering
    \includegraphics[width=\textwidth]{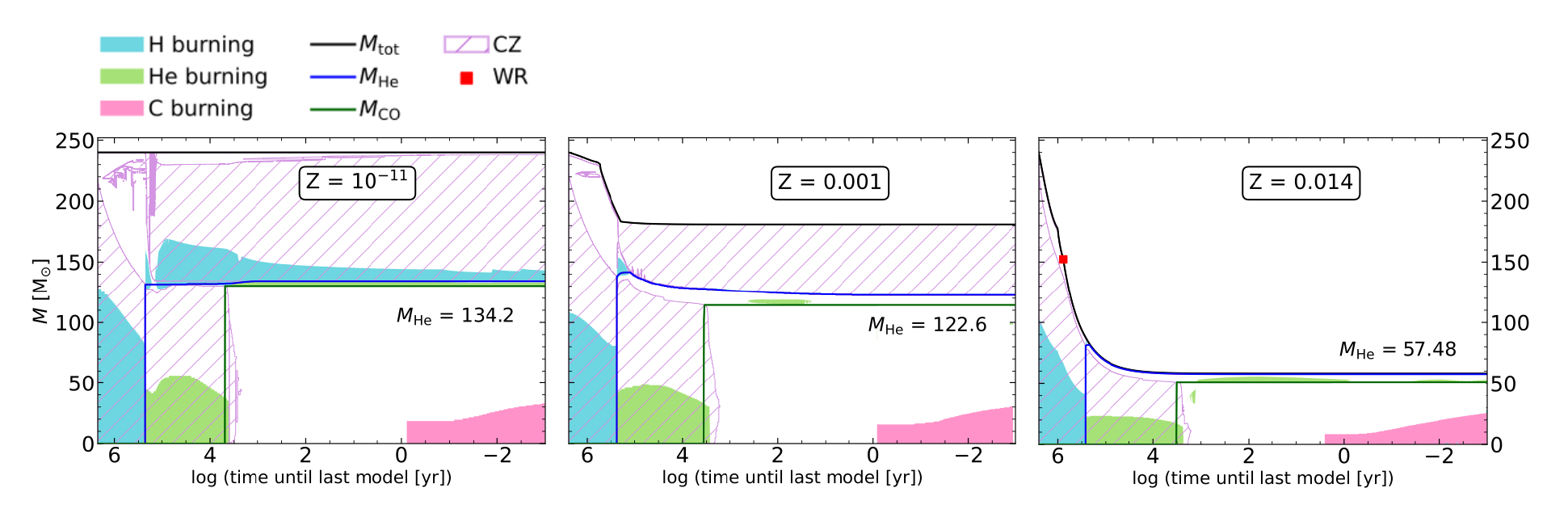}
    \caption{Kippenhahn diagrams of stars with $\Mz = 240~\msun$ with metallicity $Z = 10^{-11}$ (left), $Z = 0.001$ (center), and $Z = 0.014$ (right). 
    The black line indicates the total mass of the star.
    Blue and green lines show the He and CO cores, respectively. Burning zones are indicated in blue, green, and pink for H, He, and C, while the purple hash indicates the convective zones (CZ). The final He core is indicated on each panel. The red square indicates when $X\mathrm{^S_{H}}$ < 0.3 in mass fraction.}
    \label{fig:Kipps_joined}
\end{figure*}

\begin{figure*}
    \centering
    \includegraphics[width=\textwidth]{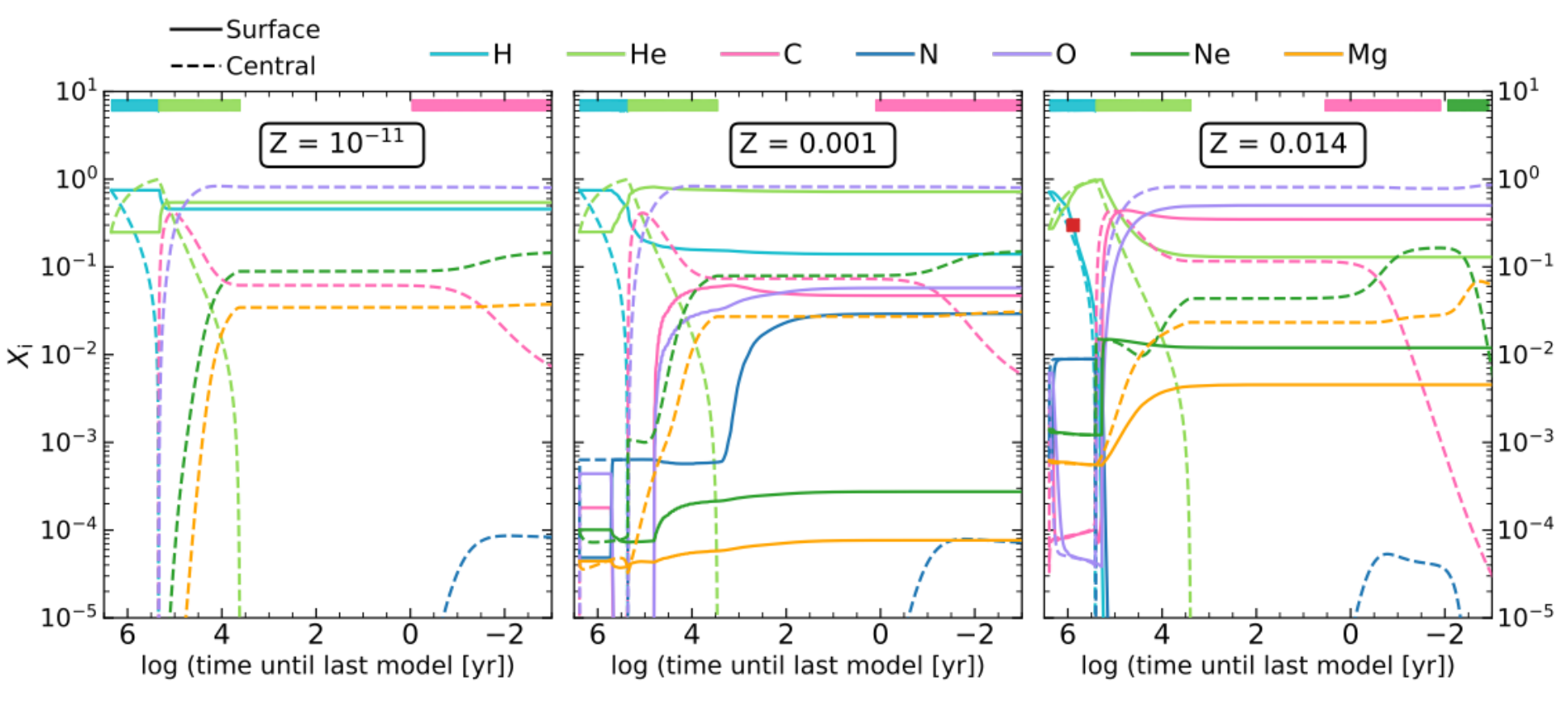}
    \caption{Central (solid) and surface (dashed) abundances as a function of time of a 240 \msun\ star for $Z = 10^{-11}$ (left), $Z = 0.001$ (center), and $Z = 0.014$(right). The abscissa is the time until the last model, and the ordinate is the mass fraction of each element. Light blue, light green, pink, dark blue, purple, dark green, and orange refer to H, $^4$He, $^{12}$C, $^{14}$N, $^{16}$O, $^{20}$Ne + $^{22}$Ne, and $^{24}$Mg+$^{25}$Mg +$^{26}$Mg abundances, respectively. Colored bars at the top indicate the type (H, He, and C) and duration of core burning phases. The red square marks the time when $X\mathrm{^S_{H}}$ < 0.3 in mass fraction.
    }
    \label{fig:abundances}
\end{figure*}

Fig.~\ref{fig:Kipps_joined} shows the Kippenhahn diagrams of stars with three selected metallicities: $Z = 10^{-11}$, $Z = 0.001$, and $Z = 0.014$, which are representative of Pop III, Pop II, and Pop I stars, respectively.
In Fig.~\ref{fig:abundances}, we show the central and surface chemical evolution of the same stellar tracks until the end of their evolution.
The Kippenhahn diagrams clearly show that the Pop III star does not lose mass via stellar winds and retains its entire mass until the pre-supernova (SN) phase.
At variance, the Pop II and Pop I stars lose much of their mass during the evolution, about 60 \msun\ and more than 180~\msun, respectively.

At the beginning of the MS, due to the high luminosity produced by the H-burning in the core, all three massive stars develop large convective regions that may involve up to 90\% of the star's total mass.
As the evolution proceeds, the convective core shrinks, and defines the size of the He core at the end of the MS.
Fig. \ref{fig:Kipps_joined} shows how, depending on the metallicity, stellar winds may peel off more of the star's external layers. In some cases, the winds are so strong that the star may become a WR already on the MS, like in the case of $Z = 0.014$.

After the MS, stars ignite He in their cores. 
Also, in this case, the high luminosity causes the core to be mostly convective.
The two models that retain their envelope after the MS (i.e., with $Z = 10^{-11}$ and $Z = 0.001$) develop large external convective regions that deepen in the stellar interior and dredge up to the surface processed material.
This dredge-up episode enriches the surface abundance of the Pop III star in He, with a mass fraction of $X\mathrm{^S_{He}} > 0.5$. 
After the dredge-up, the convective region detaches from the surface, and the chemical abundances remain constant for the rest of the evolution.
Regarding the Pop II star, the He enrichment occurs due to the concurrent effect of the mass loss stripping off the surface layers and the convective dredge up.
This makes the star's envelope H-depleted ($X\mathrm{^S_H} < 0.08$) and enriched in He, C, N, and O, at the end of the CHeB phase.
In this phase, the Pop I star's surface is quickly enriched by carbon and oxygen, and at the end of CHeB, it is essentially a bare core. 
Analogously to the end of MS, at the end of the CHeB, the size of the central convective region (inside the He core) sets the mass (and size) of the carbon-oxygen core.

After the CHeB phase, the core contracts due to neutrino losses until the central density and temperature increase enough to ignite carbon in the stellar center.
The C-burning phase lasts for a year or less, and the energy released stops the core contraction. 
From this phase on, the evolution of the core becomes so fast that it evolves decoupled from the envelope, and the stellar surface properties do not change anymore until the final stage.
At the end of the C-burning phase (or shortly after, namely, neon burning for the Pop I star), all three models become unstable to PI, and the computation is stopped, as the condition in Eq.~\ref{eq:PI} is no longer satisfied.
Although the stopping condition is similar, the pre-SN state and the final fate of the three stars are all different.

The Pop III star loses essentially no mass during its evolution and ends with a final He-core of 134 \msun. 
Due to the high mass of the core, this star undergoes a direct collapse to BH at the end of its life. 
This leads to two notable astrophysical consequences: i) the nucleosynthesis products are locked away from ever-enriching the ISM; ii) the star forms a large intermediate mass BH of $\sim$ 240 \msun. 

The Pop II star loses most of its envelope during the evolution and reaches the pre-SN phase with a total mass of about 180~\msun\ and a core mass of 122 \msun. 
The decreased core mass allows it to end its life as a PISN, ejecting all of the nucleosynthetic products into the ISM and leaving no remnant. 

Finally, the Pop I star completely loses its outer envelope during the MS, becoming firstly a WR star and then a pure He star.
The final mass of this star is 58 \msun, and it will experience pulsations driven by PI (discussed in section \ref{subsec:Fates}) before eventually collapsing into a BH. 
Due to the pulsations, about $\sim$ 13 \msun\ of material will be ejected, while the rest collapses into the BH of $\sim$ 45 \msun.
A broader discussion on the final fates of our tracks and their ejecta is given in Sec.~\ref{subsec:Fates} and Sec.~\ref{subsec:Ejecta}, respectively.

\subsection{Final fates and mass spectrum} 
\label{subsec:Fates}

\begin{figure}[]

    \includegraphics[width=0.49\textwidth]{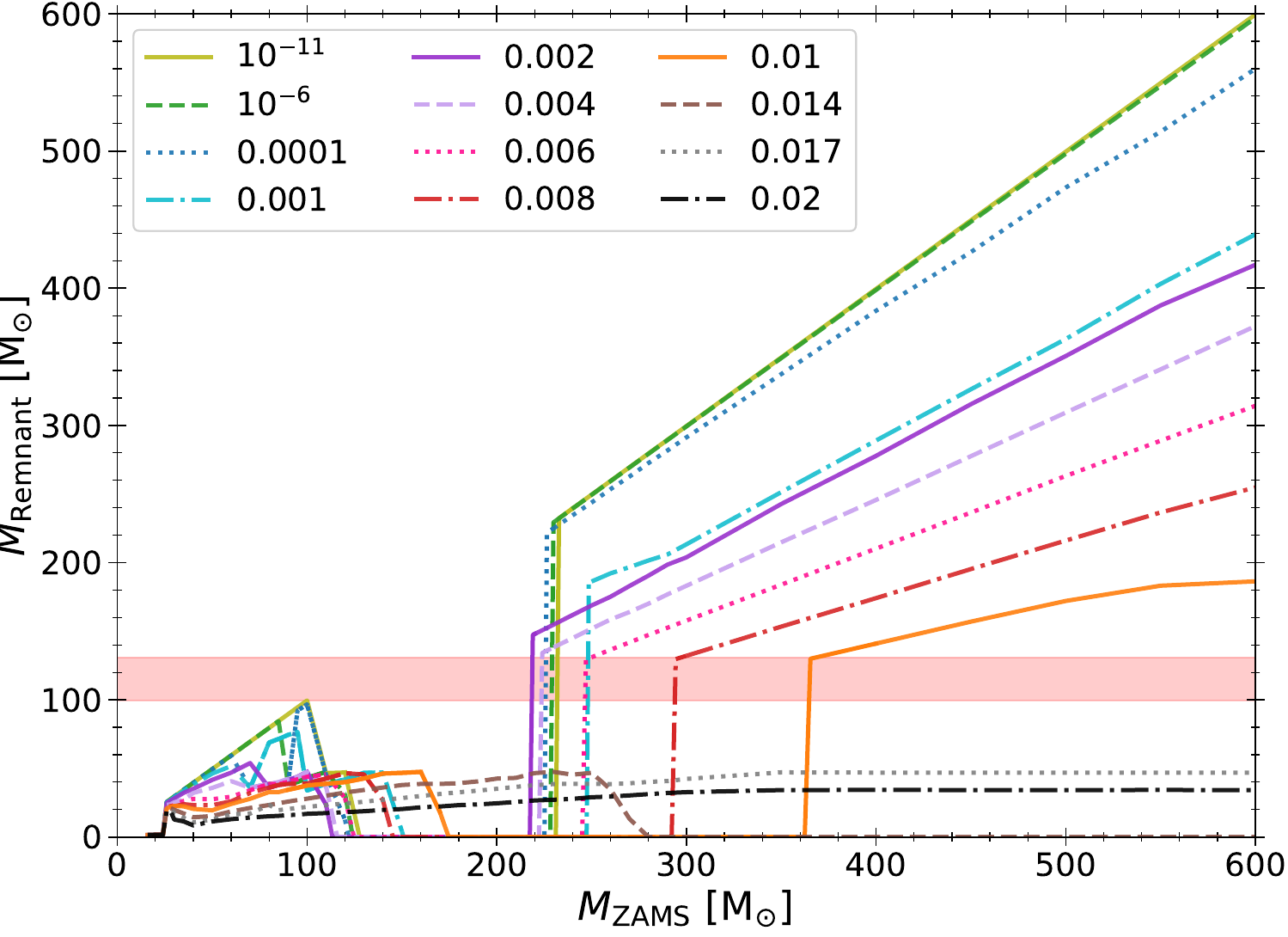}
    \caption{
    Mass spectrum of stars with different metallicities in different colors and line styles. The red-shaded area indicates the combined PI mass gap considering all metallicities that experience PI. 
    }
    \label{fig:Mass_gaps_pops}
\end{figure}

Here, we focus on determining the final fate of our massive stars with $\Mz > 14~\msun$.
We evolve such stars until the advanced stages of the oxygen-burning phase or at the onset of the pair-instability regime (see Sec.~\ref{subsec:pairs}).
The following timescales before the collapse are so short that they have negligible effects on the stars' final mass and configurations \citep{Woosley2002} and, thus, do not affect the final remnant mass estimates given in this work.

Massive stars can end their lives in many different ways. 
Depending on the pre-SN configuration, they may end up as i) core-collapse SN (CCSN), and a neutron star is left as a remnant; 
ii) failed SN (FSN), when all the envelope follows the collapse and a BH is left; 
iii) pulsational pair-instability SN (PPISN), in which the star may undergo strong mass loss episodes due to several pulses induced by PI before collapsing to a BH; 
iv) pair-instability SN (PISN), when PI triggers a single powerful pulse that totally destroys the star, leaving no remnant;
v) direct collapse into BH (DBH), when stars are massive enough that gravity wins against the PI-induced explosive ignition of oxygen, and the whole star collapses into the BH.

To obtain the final fate, we use the helium core mass ($\MHe$) and the carbon-oxygen core mass ($\MCO$) of  the last model of our tracks as a proxy to link and map the pre-supernova stage to the final fates (obtained from models by other groups).
We use the methodology described by \citet{Goswami2021} to retrieve the final fate and also the chemical ejecta (described in the section below). 
We refer the reader to their work for a detailed description of the method.
In the following, we briefly describe the main assumptions adopted to compute the final mass of the remnant and the updates on the method.
 
To account for the final phases of the stars when computing the remnant masses, we use results from other works, which model the final moments of the stars' lives with great detail.
For massive stars in the pre-SN stage with a $\MHe$ that goes from $\sim 4~\msun$ to $\sim 34~\msun$, we use results by \citet{Limongi2003} and \citet{Chieffi2004}, who modeled the final phase and the explosive nucleosynthesis of stars with $\Mz$ from $13~\msun$ to $35~\msun$, for different initial metallicities. 
These models are calibrated to reproduce the light emitted from SN1987A, and are particularly suited to obtain the nucleosynthesis of CCSN events. 
If the star in the pre-SN stage has a $\MCO$ compatible with CCSN models of \citet{Chieffi2004}, we label the star as CCSN. However, if the ejecta of $^{56}$Ni is less than $\sim$ 0.07 \msun\, or $\MCO$ is greater than any CCSN model by \citet{Chieffi2004}, the star is deemed a FSN. 
This threshold value is based on the estimated $^{56}$Ni ejected by SN1987A \citep{Nomoto2013, Prantzos2018}.
When computing the final remnant mass of stars that die as FSN, we assume that all the external envelope collapses into the newborn BH.

Following results by \citet{Woosley2017}, stars with a helium core mass between $\sim 34$ and $ 64~\msun$ end their lives as PPISN.
Stars with a \MHe\ in the range $\sim 64 - 130~\msun$ explode as PISN. 
In both cases, we require the star to be dynamically unstable (i.e., $\langle \Gamma_1 \rangle$ < 4/3 + 0.01), and is otherwise classified as FSN.
In the first case, stars experience strong mass loss episodes due to PI-triggered pulsations, which peel off the external layers before the evolution ends with a final collapse.
The remnant masses of these stars are calculated by interpolating the tabulated values from \citet{Woosley2017} based on the helium core masses in the pre-SN phase.
The explosive deaths of these very massive stars will produce an enormous amount of energy, up to $\sim$ 10$^{51}$ ergs. 

Stars of the second group, after entering the PI regime, undergo a single powerful pulse that completely destroys the star, leaving no remnant \citep{HegerWoosley2002, Heger2003}.

Stars with helium cores even greater than $\sim$ 130 \msun\ will directly collapse into BH with no explosion \citep{HegerWoosley2002}. 

During the core collapse, we consider that a certain energy is ejected in the form of neutrinos produced during the neutronization process.
The corresponding mass lost is calculated according to the results and discussion by \citet{Zevin2020}.
For FSN, PPISN, and DBH, this mass corresponds to $\sim 0.5~\msun$ while, for neutron stars, it corresponds to about $0.16 - 0.19~\msun$.

Fig.~\ref{fig:Mass_gaps_pops} shows the mass spectrum of the compact remnants -- that is,  the initial mass versus the remnant mass -- for populations of single stars with different metallicities.
The figure clearly shows the effects of two main processes that affect the massive stars' evolution: mass loss via stellar winds and the PI.
The dependency of the final BH mass on these two processes has already been largely discussed in the literature \citep{Belczynski2010, Belczynski2016, Spera2017, Iorio2023, Renzo2024}.
Due to the combined effects of winds and pair creation, each population with a different metallicity displays a distinct mass spectrum with a different mass gap in the remnant masses.
Metal-free and very-poor populations ($Z \leq 0.0001$) do not suffer from strong winds during the evolution due to their dependence on the metal content (particularly on the iron content). 
Therefore, such stars can form high-mass stellar BHs with a mass of about \Mz, if they avoid PI.
Stars with $0.0001 <  Z \leq 0.01$ suffer from increasingly stronger winds, and thus the final remnant masses are reduced. 
At $Z = 0.014$, stellar winds are very effective, and the transition between stars that undergo PPISN and PISN moves at $\sim 270$ \msun.
Above this mass limit, all stars leave no remnant due to PISN.
At this metallicity, the maximum mass BH formed after PPISN is 47.4 \msun\ from a 230 \Mz\ star. 
Above $Z = 0.014$, stellar winds become so intense that no star ends up as PISN since they lose so much mass during their evolution that the \MHe\ in the pre-SN phase is too small to trigger PI. 

\begin{figure}[]
    %\centering
    \includegraphics[width=0.495\textwidth]{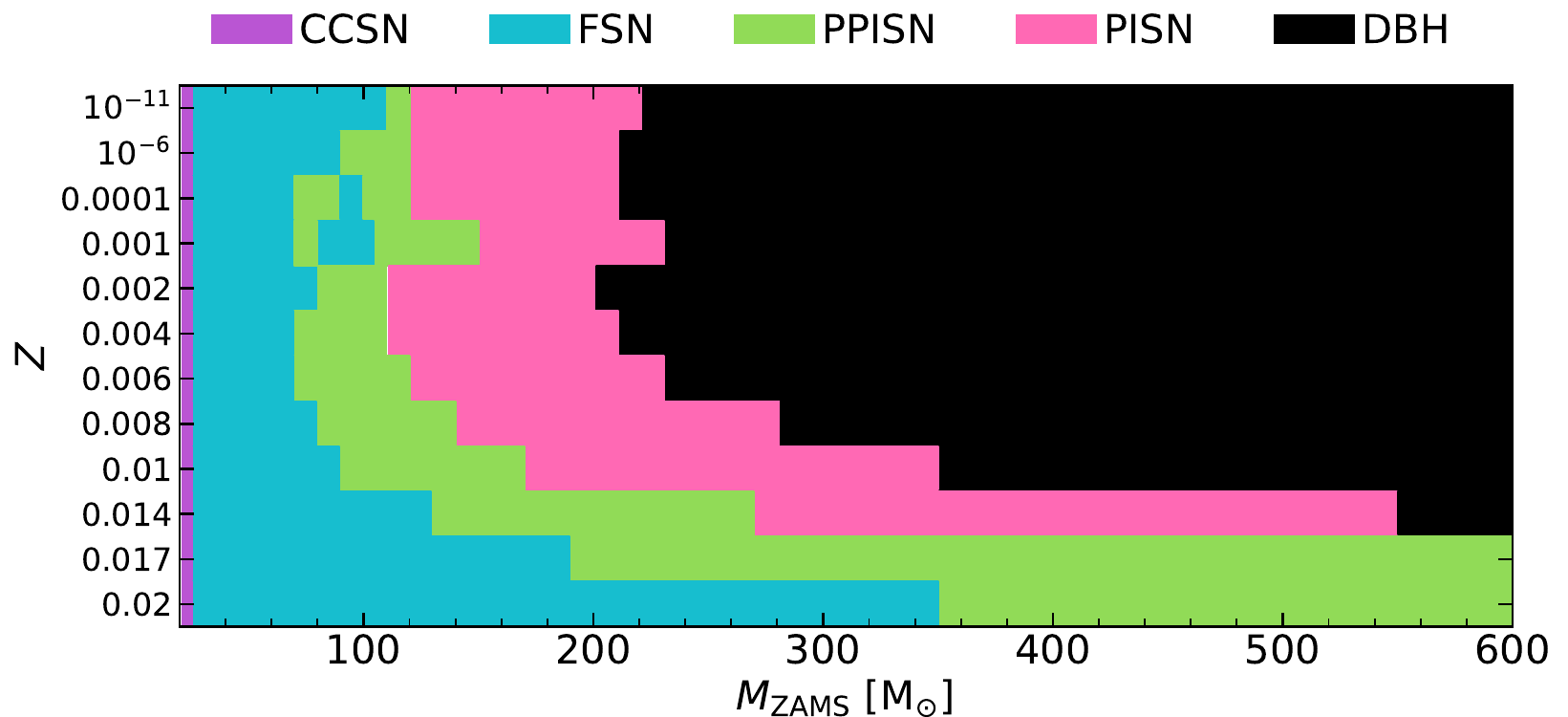}
    \caption{Final fate types for massive stars as a function of initial mass and metallicity. Purple, blue, green, pink, and black indicate the final fate of CCSN, FSN, PPISN, PISN, and DBH, respectively.
    }
    \label{fig:Remnant_type_grid}
\end{figure}

Table~\ref{tab:mgap} lists the various limits of the lower and upper edges of the mass gap for each metallicity; that is, the maximum BH mass below the PI gap and the minimum BH mass above the gap.  
To better define the initial mass of stars in the transition between PISN-DBH, we interpolate the final He cores of stars as a function of \Mz\ in the proximity of that range, and retrieve the smallest BH mass above the gap corresponding to a He core slightly above 130 \msun. 

Regardless of the initial metallicity, we find that there are no BHs predicted to form with masses $\sim$100 \msun\ to $\sim$130 \msun\ from our single star models. 
The combined PI mass gap is indicated by the red band in Fig.~\ref{fig:Mass_gaps_pops}.
We notice that this interval is smaller than PI mass gap edge predictions from other authors \citep[such as those by][]{Woosley2017, Farmer2020, Farag2022, Mehta2022} that range between $\sim 50 - 60$ and $\sim 130 - 140$ \msun, with standard nuclear reaction rates network.
This result is mainly due to the dredge-up episodes that can reduce the stellar core mass, stabilizing them against PI and thus shrinking the PI mass gap.

\begin{table}
    \caption{Mass gap edges (in solar masses) for each metallicity.}
    \begin{center}
    % \resizebox{\columnwidth}{!}{
    \begin{tabular}{c|HHc|cHc}
        \toprule
        % & \multicolumn{14}{c}{Initial composition} \\ \\
        $Z$ & $M_\mathrm{PPISN,low}$ & \Mz\ & $M_\mathrm{Gap,low}$ & $M_\mathrm{Gap,high}$ & \Mz\ & $M_\mathrm{Gap,high (interp)}$ \\
        \midrule
        $10^{-11}$ & 47.02 & 120.0 & 99.5 & 239.5 & 240.0 & 232.3 \\ 
        $10^{-6}$ & 31.7 & 120.0 & 84.4 & 229.4 & 230.0 & 229.3 \\ 
        $10^{-4}$ & 12.7 & 120.0 & 97.1 & 225.5 & 230.0  & 221.6 \\ 
        0.001 & 4.21 & 150.0 & 76.5 & 186.2 & 250.0  & 185.4 \\ 
        0.002 & 23.5 & 110.0 & 53.1 & 147.9 & 220.0 & 147.3 \\ 
        0.004 & 36.8 & 110.0 & 47.3 & 138.2 & 230.0  & 134.2 \\   
        0.006 & 28.40 & 120.0 & 45.6 & 131.0 & 250.0  & 129.4 \\ 
        0.008 & 26.8 & 140.0 & 46.5 &  132.0 & 300.0  & 129.7 \\ 
        0.01 & 18.35 & 170.0 & 47.4 & 141.1 & 400.0  & 129.9\\  
        0.014 & 13.25 & 270.0 & 47.4 & - & -  & - \\ 
        0.017 & 46.8 & 600.0 & 47.1 & - & -  & - \\ 
        0.02 & 34.1 & 600.0 & 34.3 & - & -  & - \\ 
        0.03 & - & - & - & - & -  & -\\ 
        \bottomrule
    \end{tabular}
    % }
    \end{center}

    \footnotesize{Column 1: Initial metallicity.
    Column 2: Most massive remnant below the PI mass gap. Column 3: $M_\mathrm{Remnant}$ defining the upper limit of the mass gap.  Column 4: Interpolated smallest BH mass after the PI gap.}
    \label{tab:mgap}
\end{table}

Fig.~\ref{fig:Remnant_type_grid} shows how the final fates of all the massive stars change with metallicity and initial mass, clearly showing the consequential effect of the stellar winds.
The plot also shows that the transition masses between the different fates (in particular, the one between FSNe and PPISNe) are not monotonic functions with metallicity.
Moreover, it is also clear that an island of FSNe (inside the PPISNe region) is present due to dredge-up events that lower the He core masses during the evolution of those stars, thus changing the final fate.
This non-monotonic trend of both \MHe\ and \MCO\ as a function of \Mz\ has been noticed already by \citet{Iorio2023}, using a sub-set of our tracks.
Dredge-up episodes during the CHeB phase affect not only the core mass but also the surface chemical abundances, as was shown in the previous section.
More details about this are given in Appendix~\ref{app:DUP}.

Interestingly, we find that PPISN and PISN events can occur even at and near solar metallicity, contrasting with the common assumption that PISNs only occur in very low metallicity environments \citep[as has also recently been discussed by][]{Gabrielli2024}. 

It is worth noting that the construction of the mass spectrum (with all its details) may depend on various aspects that can change its shape.
The first consists of the uncertainties related to the stellar evolution processes, which can affect the final configuration of stars in the pre-SN stage.
These include stellar winds \citep{Vink2022}, rotation \citep{Woosley2021, Volpato2024}, nuclear rates \citep{Farmer2019, Farmer2020, Costa2021}, and convective mixing processes \citep{Farrell2020, Clarkson2021, Costa2021, Volpato2023}.

The second main aspect concerns the prescriptions adopted to compute the final remnant mass. 
These are retrieved from advanced hydrodynamical studies, but the details of the final collapse are under continuous investigation \citep[][]{Fryer2012, Janka2016, Woosley2017, Fernandez2018, Patton2020, Mezzacappa2023} and, thus, the prescriptions used may be updated in the future.

\subsection{Nucleosynthetic ejecta}
\label{subsec:Ejecta}

Chemical ejecta of our stars are calculated by adopting the same methodology of \citet{Goswami2021, Goswami2022}.
We computed the chemical ejecta integrating throughout the whole evolution (for the stellar winds' ejecta) and for the different final fates (CCSN, FSN, PPISN, PISN, and DBH, explosive ejecta).
The computation includes atomic elements from H to Zn for a total of 41 isotopes\footnote{Heavier elements than those explicitly followed by \textsc{parsec} in the evolution are computed from the initial chemical composition, according to the adopted solar-scaled mixture and metallicity, because these elements are not affected by the nuclear burning phases prior to the explosive fate.}.
Winds ejecta for each different star are computed by simply integrating the ejected mass for each element from the ZAMS to the final stellar configuration.
For the explosive ejecta from CCSNe, we use and interpolate the ones by \citet{Limongi2003} and \citet{Chieffi2004} to include the newly synthesized elements during the SN event \citep[for more details, we refer the reader to][]{Goswami2021}.
For models that finish their evolution with an FSN or a DBH collapse, no explosive ejecta are produced \citep{Fryer2001, Heger2002, Nomoto2013}, and the only chemical contribution to ejecta comes from stellar winds.
Models that die as PPISNe eject a large amount of mass during the pre-collapse pulsations, which are accounted for in the ejecta.
The explosive nucleosynthesis and the amount of mass lost due to pulsations are computed by interpolating results from \cite{Woosley2017} as a function of the pre-SN helium core mass. 
After the pulsational phase, the star collapses into a BH, and no further ejecta are produced.
Finally, for stars that explode as PISNe, we computed the explosive ejecta using the results by \citet{HegerWoosley2002}, using the mass of the helium core as a proxy.

The chemical ejecta tables present the components derived from both stellar winds and the combined total, containing the winds and the explosive ejecta, and are also available in the web database\footnote{\url{http://stev.oapd.inaf.it/PARSEC/}}.

\begin{figure}[h!]
    \centering
    \includegraphics[width=\columnwidth]{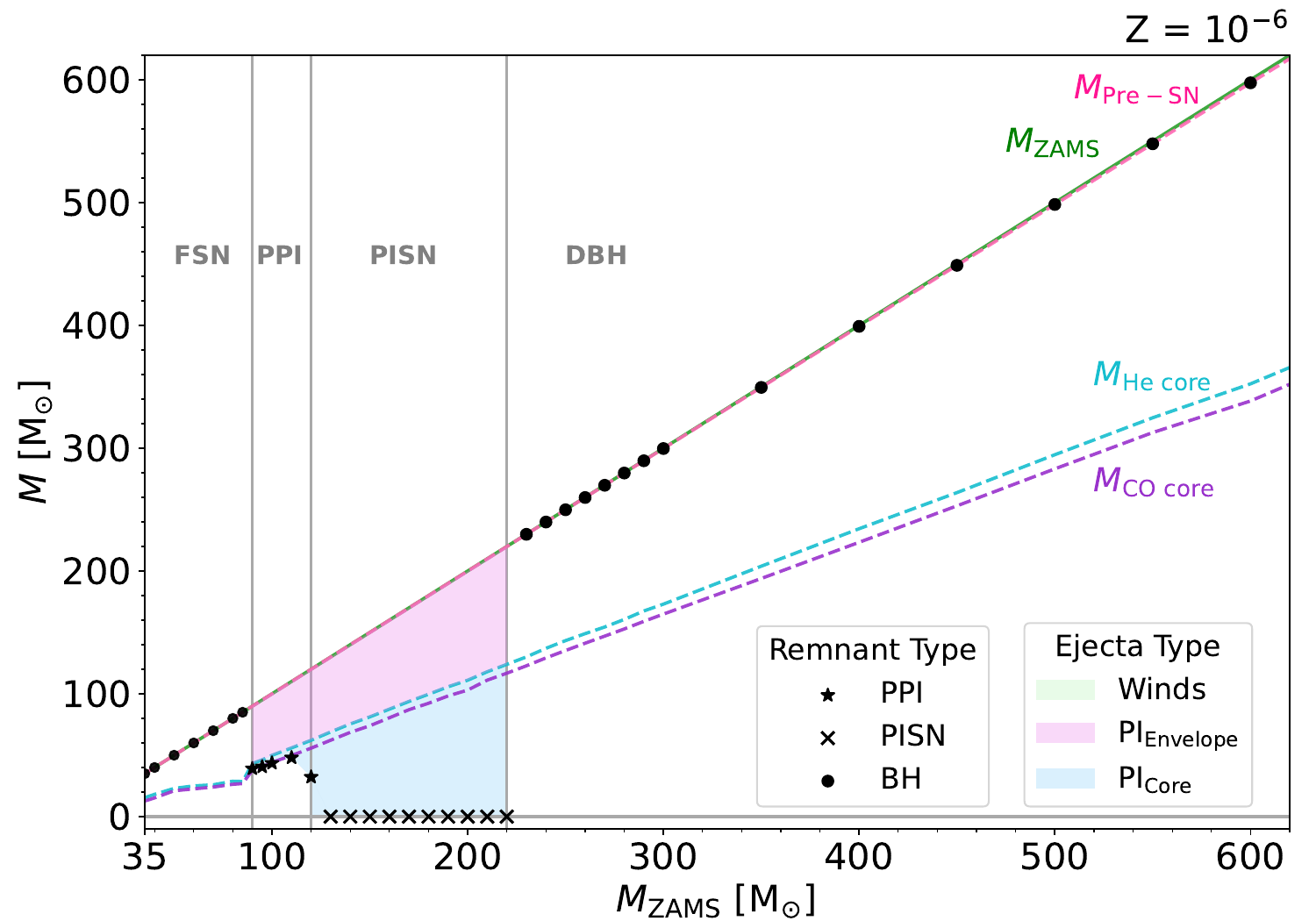}
    \includegraphics[width=\columnwidth]{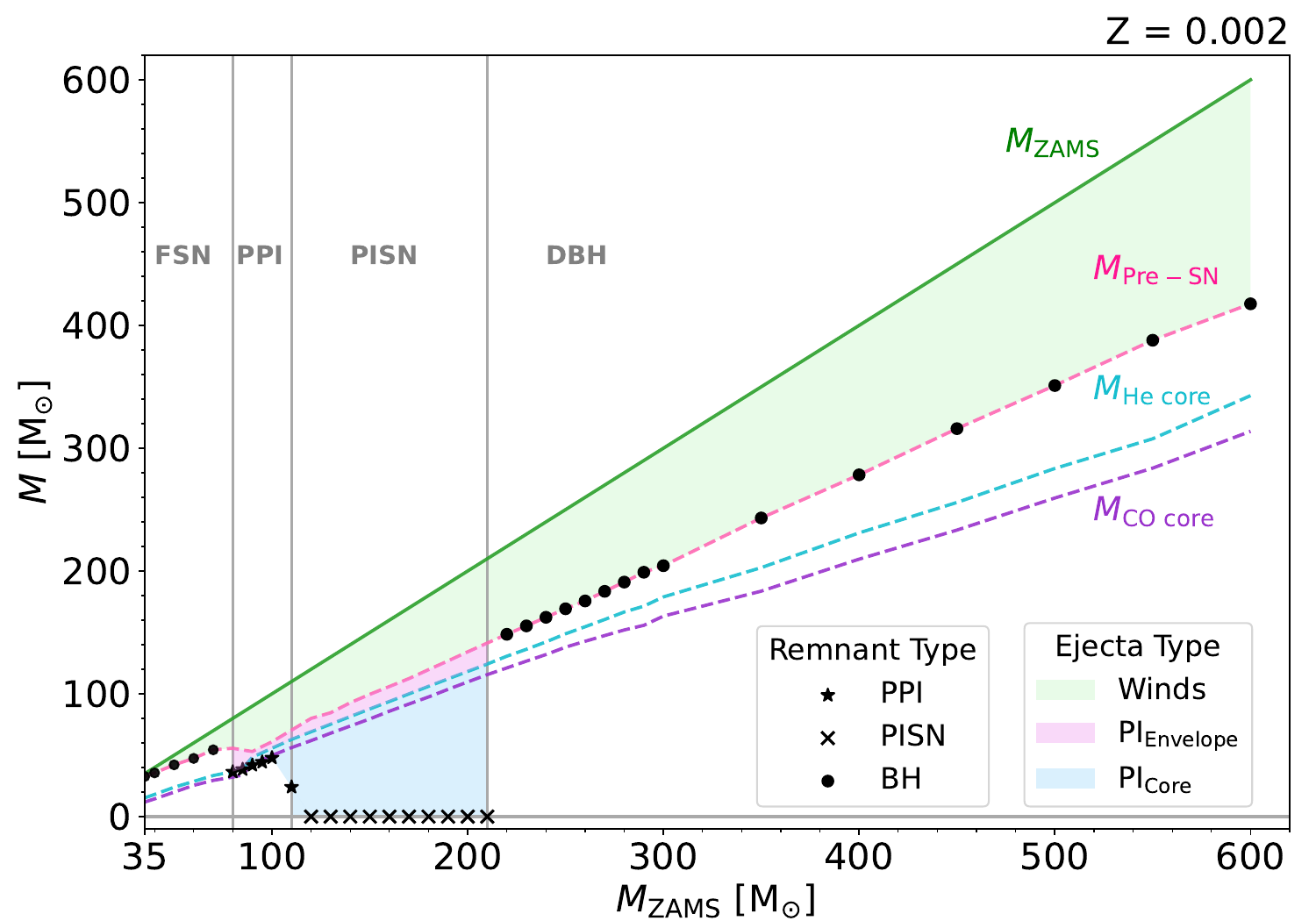}
    \includegraphics[width=\columnwidth]{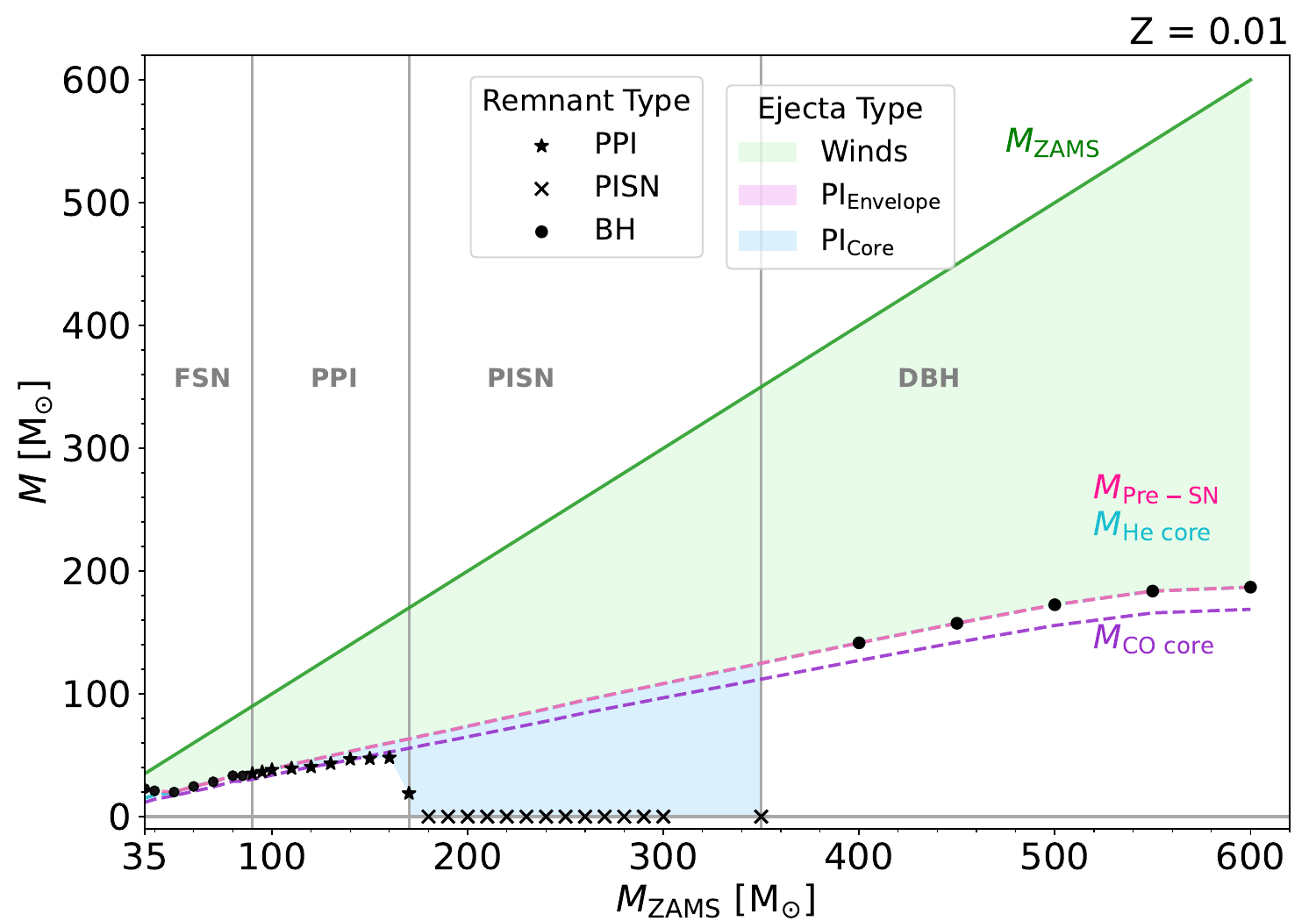}
    \caption{Remnant mass as a function of the initial mass for tracks with $Z = 10^{-6}$, $Z = 0.002$, $Z = 0.01$, in the top, middle, and bottom panels, respectively. 
    The solid green line indicates the initial mass. The dashed pink, blue, and purple lines indicate the final mass of the star, the He core mass, and CO core mass, respectively. Black markers indicate the different fate of the models.
    The shaded green area indicates the mass lost during the evolution due to stellar winds. The shaded pink and blue regions are the outer envelope mass and the core mass ejected during the final fate, respectively.
    }
    \label{fig:tot_ejecta}
\end{figure}

An example of the various origins of the ejecta from tracks with different metallicity can be appreciated in Figure \ref{fig:tot_ejecta}.
As has been discussed before, low metallicity stars lose very little mass during their evolution because stellar winds are highly metallicity-dependent.
Therefore, these stars can form the largest stellar remnants, which, in the case of FSN or DBH, have a mass very close to the initial one.
At these low metallicities, ejecta are mainly due to the explosive CCSNe, PPISNe, and PISNe fates.
As the metallicity increases, the winds' contribution to the total ejecta increases, becoming the most significant one (in solar masses) for stars with $Z > 0.001$.
Depending on the final fate, stars eject material from different layers.
For instance, during the pulsational phase of a PPISN, stars generally eject material from the envelope well above the He core, with the exception of the most massive ones, which experience powerful pulsations that can extract material even from the core.
On the other hand, CCSNe or PISNe always eject the highly enriched material from the stellar core.

\begin{figure}
    \centering
    \includegraphics[width=\columnwidth]{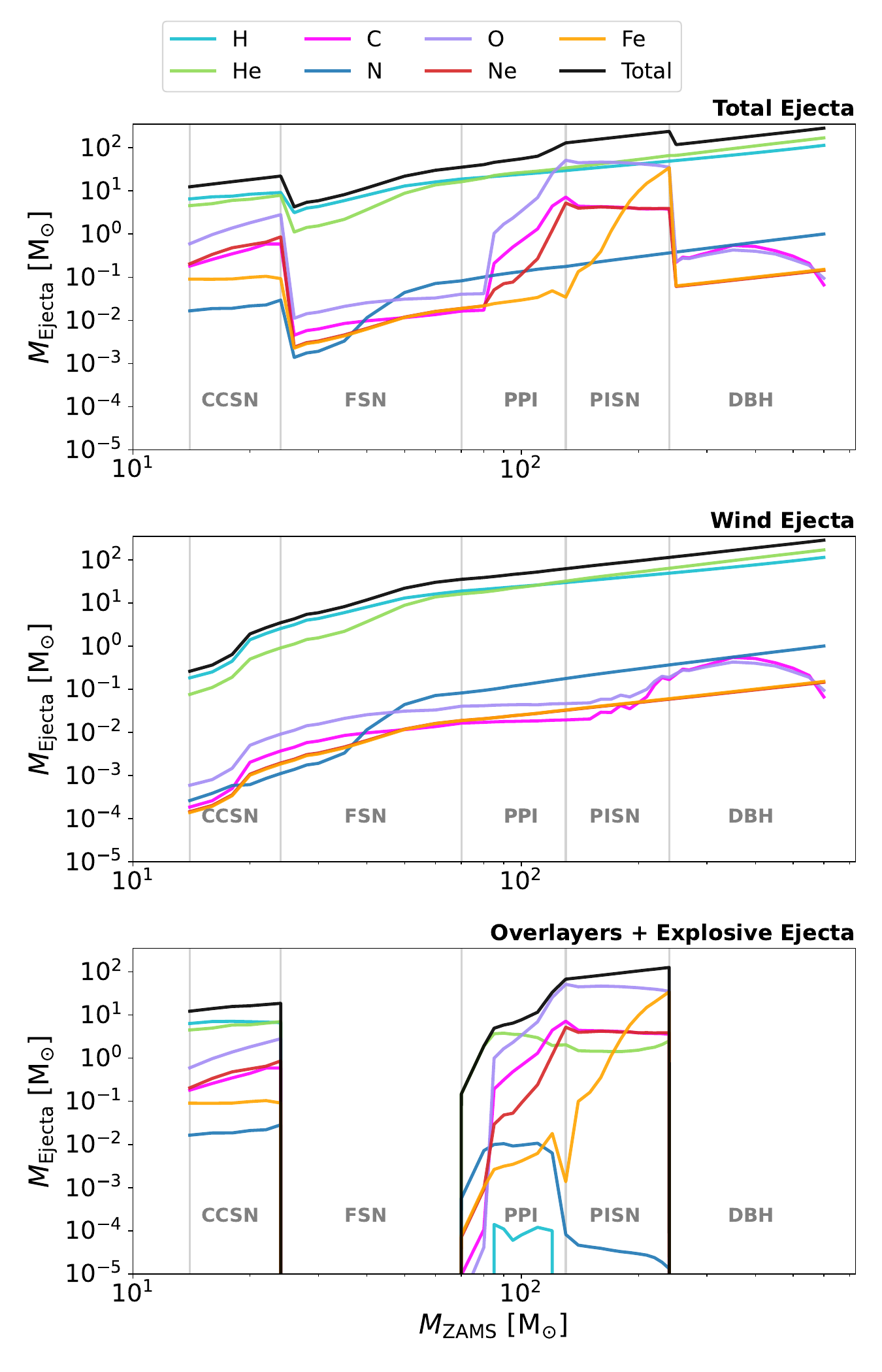}
    \caption{Chemical makeup of ejecta for $Z = 0.006$ stars.  
    The top panel indicates the total mass lost, the middle is the mass loss due to winds, and the bottom panel shows the sum of the elements from the inner layers ejected (overlayers) plus the explosive ejecta of stars that enter pair-instability or undergo core collapse. The total ejected mass is indicated by the solid black lines, whereas the ejecta in the form of H, $^4$He, $^{12}$C, $^{14}$N, $^{16}$O, $^{20}$Ne, and Fe are marked by cyan, light green, pink, navy, purple, red, and orange lines, respectively. 
    Fe refers to the elemental Fe and the isotopes $^{52}$Fe through $^{51}$Fe. 
    Vertical gray lines separate the regimes of stars that enter CCSN (14 $\leq$ \Mz/\msun\ $\leq 24$), FSN (26 $\leq$ \Mz/\msun\ $\leq 60$),  PPI (70 $\leq$ \Mz/ \msun\ $\leq 120$), PISN (130 $\leq$ \Mz/\msun\ $\leq$ 240), and DBH (\Mz\ $\geq 250$ \msun), for this metallicity.
    } 
    \label{fig:total_ej_abund_z0.001}
\end{figure}

It is interesting to evaluate the chemical makeup of the ejecta of massive stars and see which ones (and their initial masses) contribute the most to the enrichment of the interstellar medium. 
Fig.~\ref{fig:total_ej_abund_z0.001} shows the contribution of each type of ejecta for seven of the most abundant elements for the sets of tracks computed with $Z = 0.006$.
Ejecta generally increase with the initial masses, but the differing final fates impact the trend and the detailed abundance outcome.
Stars with a mass $14 \le \Mz/\msun \le 24$ end their evolution as CCSNe. 
Their ejecta are rich in elements produced during the evolution (like He and C) and from the explosion (mostly Fe), while the contribution from stellar winds is still low. 
The total ejecta decreases as stars die as FSNe since all the mass collapses in the final BH, and the winds are the only contributors.
Despite the final collapse into a BH, stars with \Mz\ $\geq 50$ \msun\ eject tens of solar masses as winds start to be very efficient.
Moving up in the mass regime, stars with $70 \leq \Mz/\msun \leq 130$ enter the PPISN regime, and the ejected mass increases.
In this mass regime, %\textcolor{red}{explosive} 
ejecta are strongly enriched by 
elements processed during evolution, like carbon, oxygen, and neon, in the stars' inner layers (also called overlayers). 
As the \Mz\ increases, the PI-driven pulsations' energy increases, and more matter is expelled, leading to an increase in the ejected matter.
The total ejecta reach their peak for stars that experience PISN ($130 \leq \Mz/\msun \leq 240$).
The ejecta from these stars are strongly enriched by Fe (and $^{56}$Ni, which is included in the Fe ejecta) produced in the explosion, which increases with the initial mass.
Conversely, the amount of carbon, oxygen, and neon ejected is nearly constant with \Mz\ for these stars. 
At the same time, the total amount of H, He, and N are nearly unchanged by the explosion, and their contribution to the total ejecta comes almost entirely from the winds.
Finally, stars with $\Mz \geq 240~\msun$ undergo DBH collapse. 
Therefore, their ejecta are only due to stellar winds, which are primarily composed of H and He, as the outermost layers of the star are ejected.
We also stress that stars that do experience strong dredge-up episodes may show an increase in the products of the H-burning since it brings fresh hydrogen to the H-burning shell and takes processed elements up to the surface (see App.~\ref{app:DUP} for more details).

\begin{figure}[]
    \centering
    \includegraphics[width=\columnwidth]{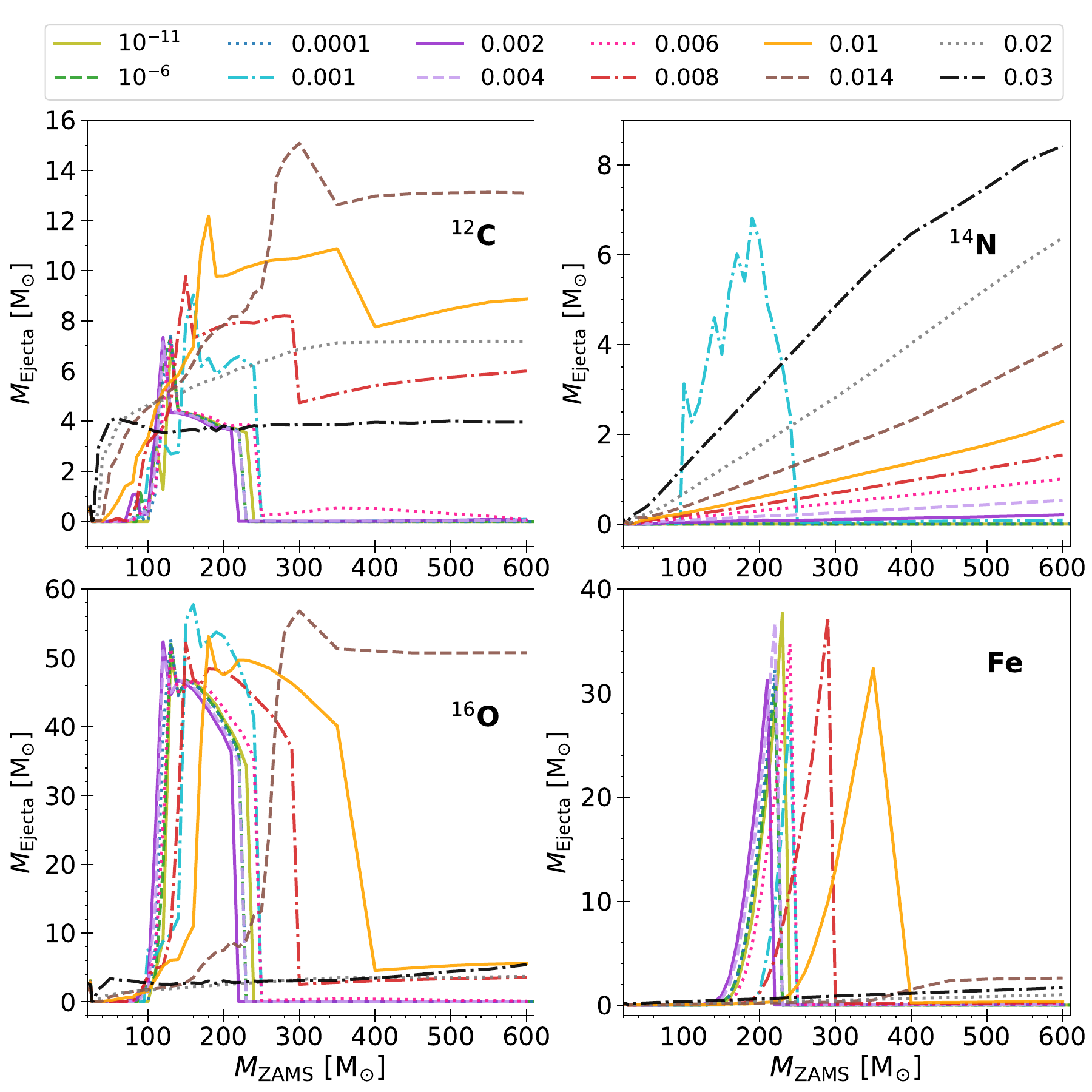}
    \caption{Total amount of mass ejected for 4 species. Fe refers to the isotopes $^{52}$Fe through $^{51}$Fe and the $^{56}$Ni that decays into $^{56}$Fe.
    It is worth noting the peculiar behavior of $^{14}$N for $Z = 0.001$, resulting from the specific interplay between the dredge-up history and the final fate of these tracks (see text for more details).
    }  
    \label{fig:M_ini_Mejecta_4}
\end{figure}

To show the relative difference in the ejecta between various metallicities, Fig.~\ref{fig:M_ini_Mejecta_4} shows the total ejecta of four elements at various metallicities.
The figure shows a clear trend with the metallicity.
All stars with metallicities up to 0.006 show similar but increasing ejecta as the metallicity increases.
The only exception that does not follow such a trend is the set with $Z = 0.001$, as is shown in the panel of the total N ejecta.
For this metallicity, the high ejecta of Nitrogen in the mass range $100 \leq \Mz/ \msun \leq 240$ are almost entirely due to the strong dredge-up experienced during and after the CHeB phase.
This strong dredge-up is already present at $\Mz=80~\msun$ (see Fig.~\ref{fig:app_kipp_Z0.001}), but in this case, nitrogen is not ejected because the star ends as a FSN.
On the contrary, in the higher mass range, all the nitrogen produced due to dredge-up is ejected due to pulsations or to the final PISN explosion (this is the case of the $\Mz = 240~\msun$ star shown in the middle panel of Fig.~\ref{fig:Kipps_joined}).

Ejecta of more metallic stars ($Z > 0.006$) show a more pronounced growth (with $Z$), mainly due to the effect of winds.
Furthermore, we stress that the large amount of carbon and, in particular, of oxygen ejected by stars with masses $\Mz> 300~\msun$ at $Z = 0.014$ is due to their final explosion as PISN.
We note that the high [O/Fe] abundance ratio contribution from PISNs at low metallicity has been invoked as a possible way to reproduce the chemical evolutionary path observed in halo and thick-disk stars in the Galaxy \citep[see][]{Goswami2021}, in alternative to more conventional interpretations \citep{Micali2013, Grisoni2017, Matteucci2021}.
When the ZAMS mass approaches the upper limit for the PISN phase, the contribution of Fe becomes very high, even at very low $Z$ (see bottom right panel of Fig.~\ref{fig:M_ini_Mejecta_4}).
These stars could be responsible for the high [Fe/O] ratio observed in extremely metal-poor galaxies \citep[see][]{Izotov2018, Kojima2020, Goswami2022}.

In addition to what has been presented until now, it is important to mention that other important physical processes can significantly impact the final total ejecta, especially stellar rotation \citep{Hirschi2005,  Ekstrom2008, Limongi2018, Goswami2022}.
Rotation can strongly influence stellar nucleosynthesis by inducing extra mixing in the stable layers of the star. 
This extra mixing leads to differences in the nuclear burning lifetimes, chemical abundances of the star, final fates, and yields. 
This is particularly important for the nitrogen enrichment of the interstellar medium \citep{Meynet2006}.
Studying the effect of rotation in our models is out of the scope of this paper, and will be investigated in future works.

\begin{figure*} % This is updated with new photons 
    \centering
    \includegraphics[width=\textwidth]{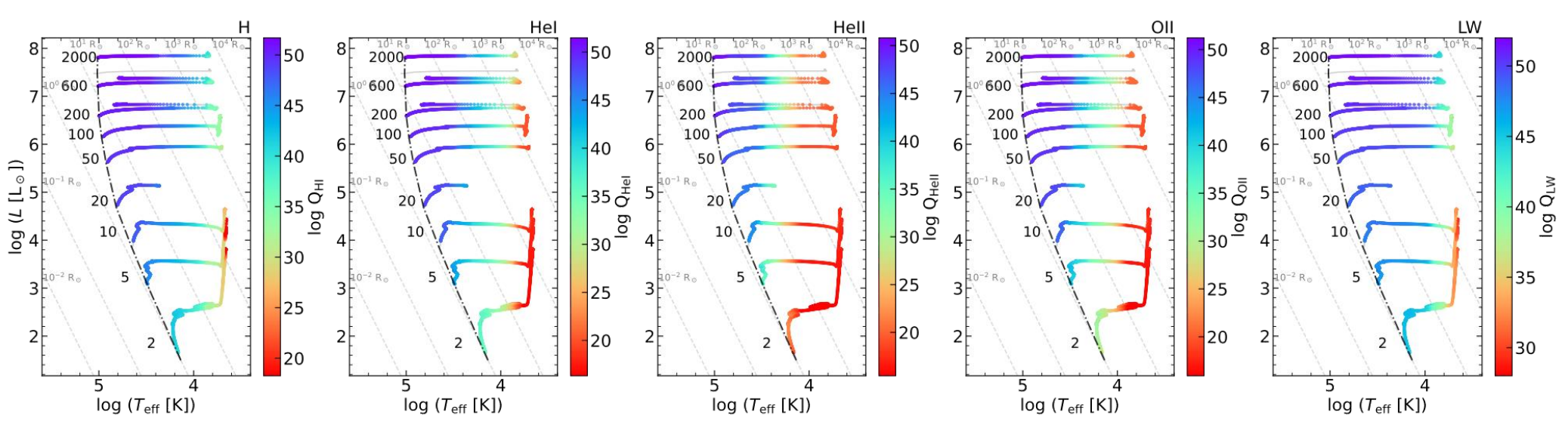}
    \caption{HR diagrams of selected $Z = 10^{-11}$ tracks, color-coded by the log number of ionizing photons emitted per second. From left to right, the panels show the log number of ionizing photons of each type emitted per second ($Q_\mathrm{HI}$, $Q_\mathrm{HeI}$, $Q_\mathrm{HeII}$, $Q_\mathrm{OII}$, and $Q_\mathrm{LW}$). Tracks are labeled by their initial mass in solar units.}
    \label{Z1D-11_ionizing}
\end{figure*}

\begin{figure*}
    \centering
    \includegraphics[width=\textwidth]{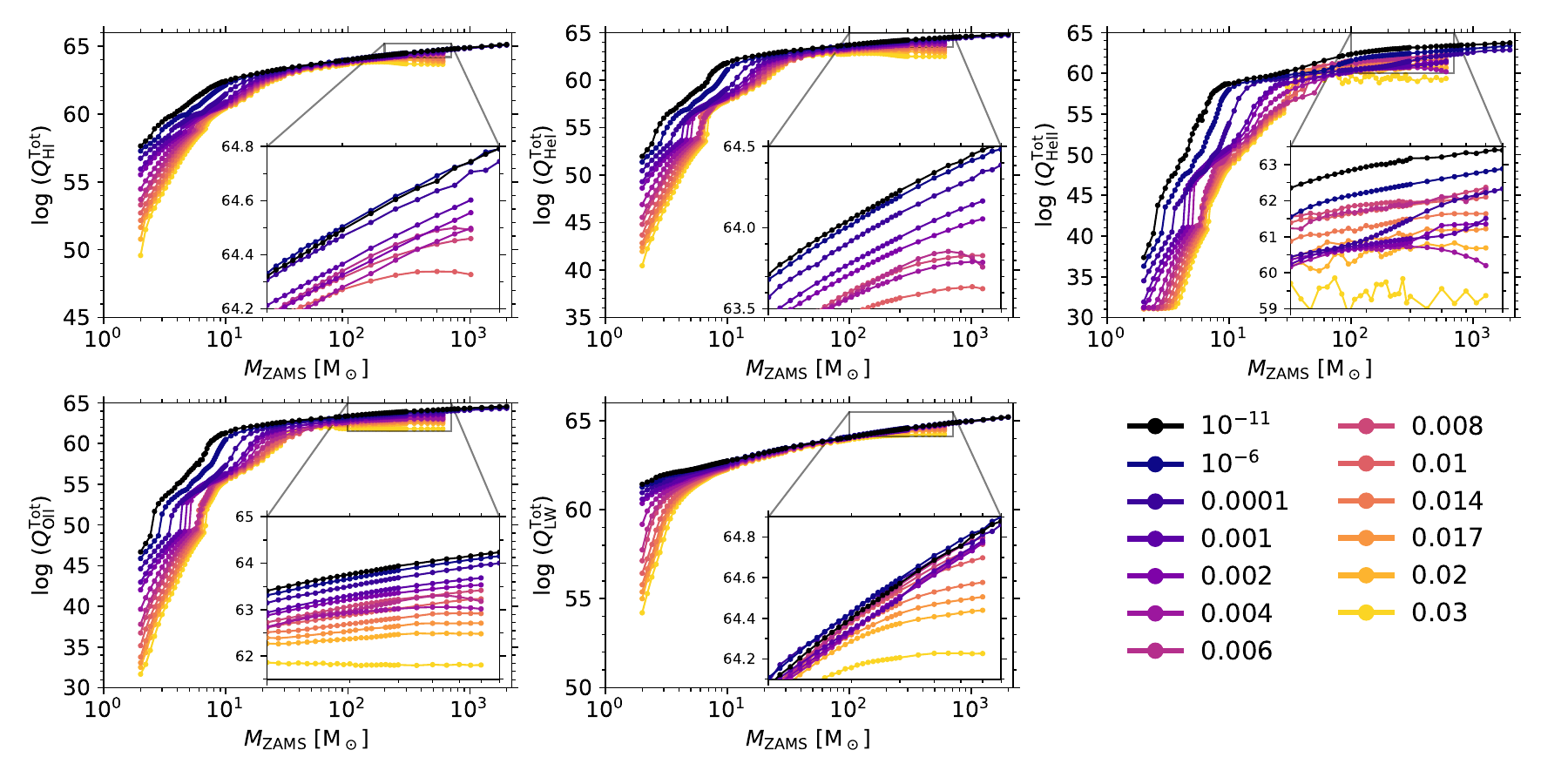}
    \caption{Total number of ionizing photons emitted throughout a star's lifetime as a function of its initial mass and metallicity. Different panels indicate the distinct ionization types considered here.}
    \label{fig:ionizing_phot}
\end{figure*}

\subsection{Ionizing radiation}
\label{subsec:Ioniz_rad}

Massive ($10 \leq \Mz/\msun \leq 100$) and very massive ($\Mz \geq 100~\msun$) stars are a major source of ionizing radiation in the Universe \citep{Topping2015, Martins2023b, Schaerer2024}. 
Here, we explore the trend with metallicity of the ionization photons released by these stars.

We computed the number of photons per second (the production rate) emitted from the star's surface with energies high enough to ionize H (HI, Lyman continua), He (HeI), He$^+$ (HeII), O$^+$ (OII), and the UV Lyman-Werner (11.2 $-$ 13.6 eV) band, able to dissociate H$_2$.
To perform the calculations, we adopted the stellar spectra database from \citet{Chen2015}, later updated by \citet{Chen2019}, in which also bolometric correction tables have been included and released in the \textsc{YBC} database\footnote{\url{http://stev.oapd.inaf.it/YBC/}}.
This collection includes spectra computed with the \textsc{atlas9} \citep{Kurucz2014}, \textsc{phoenix} \citep{Allard2012}, \textsc{WM-basic} \citep{Pauldrach1986} and \textsc{PoWR} \citep{Todt2015} codes.
We calculate the ionizing photon production rates in these five bands for all stars in our sets, using the current mass, stellar radius, effective temperature, and surface abundance during the evolution. 
The tables of ionizing photons are available in the \textsc{parsec} tracks database\footnote{\url{http://stev.oapd.inaf.it/PARSEC/}}.

Fig. \ref{Z1D-11_ionizing} shows the HR diagram of $Z = 10^{-11}$ colored by the number of ionizing photons emitted per second ($Q_\mathrm{HI}$, $Q_\mathrm{HeI}$, $Q_\mathrm{HeII}$, $Q_\mathrm{OII}$\footnote{O$^+$ ionizing photon rates, combined to the others, are useful to retrieve other spectral lines from fitting formulae by \citet{Panuzzo2003}.}, $Q_\mathrm{LW}$). 
As was expected, the ionizing rates increase with the effective temperature and luminosity.
Therefore, massive stars have higher ionizing production rates, and during their evolution, they emit much more ionizing radiation when on the blue side of the diagram, mostly during the MS.
As stars evolve to the red side, they produce fewer ionizing photons, but the drop happens at different effective temperatures depending on the band.
Massive stars with $\Mz \geq 50~\msun$ can be strong ionizing, and dissociating, sources for H and H$_2$ still at $\teff \sim 10^4$ K.
While, for He, He$^+$ and O$^+$, they lose much of their ionizing strength when $\teff \leq 1.6\times10^4$ K.
This general trend is found for all metallicities.
At higher metallicities, the appearance of WR stars due to stellar winds changes the star's path in the HR diagram, favoring stars on the blue side.
This also increases the ionizing photon production rate in the post-MS phases.
This trend for Pop III and for stars with higher metallicity is found similar to that of other authors in the literature \citep[for instance see][]{Topping2015, Murphy2021, Klessen2023}.

Two ingredients are needed to understand how much single stars contribute to the overall production of ionizing photons: the stellar surface properties and lifetimes.
To study how the total ionizing photon production varies with metallicity, we computed the lifetime-integrated values from the rates ($Q_{i}$).
Hence, $Q^\mathrm{Tot}_{i}$ are the total number of ionizing photons emitted during the whole star's lifetime (starting from ZAMS).
Here, we neglect the contribution to the total emission of ionizing photons due to the final fates of the stars.
Fig.~\ref{fig:ionizing_phot} shows the $\log Q^\mathrm{Tot}_{i}$ vs. $\log \Mz$ for different initial metallicities. 
In the \Mz\ mass range between 2 \msun\ and 30 \msun, the total number of ionizing photons shows a broad excursion (several orders of magnitude) for different metallicities (the higher the Z, the lower the $Q^\mathrm{Tot}_{i}$) due to the hotter MS of low-$Z$ stars (see Fig.~\ref{fig:app_ZAMS}).
In the range of masses with $\Mz > 30 \msun$, $Q^\mathrm{Tot}_{i}$ still increases with mass, but the dependence on metallicity becomes weaker.
At such a high mass regime, the difference due to the ZAMS position is quenched by the effect of stellar winds at higher metallicities since they increase the time spent on the blue side of the HR by the star. 
Thus, the total contribution to the ionizing photon production of these stars increases, too.

An interesting feature shown in the figure is that Pop III stars ($Z = 10^{-11}$) do not always produce the highest total number of ionizing photons for H, and dissociating for H$_2$.
The zoom-in panels show that for $\Mz > 100~\msun$, the integrating number of photons is slightly higher for stars with $Z = 10^{-6}$.
This is because, above certain effective temperatures, the production rates of ionizing photons do not increase much with \teff.
In this case, even though Pop III stars are hotter in MS (see the ZAMS in the left panel of Fig.~\ref{fig:app_ZAMS}), the averaged rates on the lifetimes are very similar for stars with $Z = 10^{-6}$ and $10^{-11}$ (as the right panel of Fig.~\ref{fig:app_ZAMS} shows).
Since stars with $Z = 10^{-6}$ have longer lifetimes (Fig.~\ref{fig:app_lifetimes}), the total number of photons could be slightly higher.
This result suggests that, as Pop III stars, also very metal-poor Pop II stars can be strong UV emitters, which could ionize the hydrogen atoms and dissociate the H$_2$ molecule, thus similarly impacting the environment.
Moreover, considering that strong UV feedback is one of the key ingredients that lead to a top-heavy IMF for Pop III stars \citep[see][for a review]{Klessen2023}, also very metal-poor stars could have a strong UV feedback, and so, a top-heavy IMF \citep[for instance, see new results by][]{Chon2021, Chon2024}.

Computations of ionizing photons are at the base of more detailed studies on integrated stellar populations in clusters and galaxies \citep[e.g., see][]{Schaerer2024}.
In this context, binaries play an important role, as they can produce large differences in the amount of ionizing photons from a certain population with respect to populations of single stars.
This is because binary evolution can produce more stripped blue stars (namely strong UV emitters), which cannot be formed via standard single-star evolution.
Recently, studies by \citet{Stanway2018} and \citet{Lecroq2024} show that binaries are necessary to correctly interpret and reproduce the emission of young metal-poor star-forming galaxies.

It is worth saying that there is still a certain amount of uncertainty regarding the rate of ionizing photons from models.
For instance, by exploring the uncertainties in the mixing processes, \citet{Murphy2021} find that small variations in the internal H-proﬁle can affect the rate of ionizing photons by up to 30\% at the end of the MS.
Other comparisons by \citet{Agrawal2022} show that the Lyman flux emitted by a synthetic population with $Z = 0.014$ from stellar tracks produced by different evolutionary codes can differ as much as $\sim 18\%$.

%-----------------------------------------------------------------

\section{Discussion}
\label{sec:discussion}

In this section, we compare our new standard sets of stellar tracks with other codes and with observations. 
More detailed descriptions of several other properties of our tracks are given in the Appendix.

\subsection{Comparison with other codes}
\label{subsec:comp_w_other_codes}

Here, we compare our stellar tracks with results from other authors.
In particular, with \textsc{mist} tracks \citep{Choi2016} from the \textsc{mesa} code \cite{Paxton2011}, with 
models by \citet{Ekstrom2012} computed with the Geneva code \citep[\textsc{genec},][]{Eggenberger2008}, and with tracks from the \textsc{franec} code \citep{Chieffi2013, Limongi2018}. 
The main differences in the stellar input parameters (for convection treatment and overshooting) are summarized in Table \ref{tab:code_comparison}. 
Moreover, there are different wind prescription choices and treatments for the density inversion experienced by RSG massive stars (see Sec.~\ref{subsec:micro_phys} for a description of our approach).
For a more in-depth review of the differences between the massive stars models computed with these codes (and others), we refer to \citet{Agrawal2022} and \citet{Martins2013}.

\begin{table}
    \caption{Input details of tracks from different codes shown in Fig.~\ref{fig:HRD_Comparison}.}
    \begin{center}
    % \resizebox{\columnwidth}{!}{
    \begin{tabular}{c|cccc}
        \toprule
        Code & $Z$ & Convection & $\alpha_\mathrm{MLT}$ & $l_\mathrm{ov}$ \\
        \midrule
        \textsc{parsec} & 0.014 & Schw. & 1.74 & 0.25 \\ 
        \textsc{genec} & 0.014 & Schw. & 1.6 & 0.1 (step) \\ 
        \textsc{mist} & 0.01428 & Ledoux & 1.82 & 0.016 (exp.) \\ 
        \textsc{franec} & 0.01428 & Ledoux & 2.3 & 0.2 \\ 
        \bottomrule
    \end{tabular}
    \end{center}
    % }
    \footnotesize{Column 1: stellar code name. Column 2: metallicity. Column 3: adopted convection criterion. Column 4: mixing length in $H_P$. Column 5: overshooting distance in $H_P$.}
    \label{tab:code_comparison}
\end{table}

\begin{table}
    \caption{Final properties of tracks from different codes shown in Fig.~\ref{fig:HRD_Comparison}.}
    \begin{center}
    \resizebox{\columnwidth}{!}{
    \begin{tabular}{c|cccc}
        \toprule
        \Mz\ [\msun] & $M_\mathrm{Pre-SN}$ [\msun]& $M_\mathrm{He}$ [\msun]& $\log (\teff [\mathrm{K}])$ & $\log (L_\mathrm{pre-SN}~[\mathrm{L}_\odot])$ \\
        \midrule
        \multicolumn{5}{c}{\textsc{parsec}}\\
        \midrule
        10 & 8.76 & 2.72 & 3.54 & 4.57 \\
        20 & 18.37 & 7.126 & 3.53 & 5.28 \\
        40 & 14.86 & 14.70 & 5.40 & 5.80 \\
        120 & 32.94 & 32.60 & 5.31 & 6.27 \\
        300  & 69.70 & 69.00 & 3.91 & 6.64 \\
        \midrule
        \multicolumn{5}{c}{\textsc{genec}}\\
        \hline
        10 & 9.68 & 2.53 & 3.55 & 4.46 \\
        20 & 8.63 & 3.65 & 3.57 & 5.18 \\
        40 & 12.82 & 10.37 & 3.85 & 5.67 \\
        120 & 30.91 & 25.48 & 4.80 & 6.25 \\
        300 & 38.15 & 38.15 & 4.26 & 6.37 \\
        \midrule
        \multicolumn{5}{c}{\textsc{mist}}\\
        \hline
        10 & 9.41 & 2.74 & 3.52 & 4.63 \\
        20 & 14.09 & 6.57 & 3.56 & 5.23 \\
        40 & 23.30 & 15.94 & 4.02 & 5.79 \\
        120 & 11.86 & 11.86 & 5.31 & 5.60 \\
        300 & 15.67 & 15.66 & 5.37 & 5.80 \\
        \midrule
        \multicolumn{5}{c}{\textsc{franec}}\\
        \hline
        20 & 7.54 & 3.86 & 4.00 & 5.23 \\
        40 & 14.14 & 10.58 & 5.27 & 5.70 \\
        120 & 27.87 & 21.95 & 5.36 & 6.16 \\
        \bottomrule
    \end{tabular}
    }
    \end{center}
    \footnotesize{Column 1: Initial mass. Column 2: pre-SN mass. Column 3: He-core mass for the final. Column 4: Final effective temperature. Column 5: final luminosity.}
    \label{tab:code_comp_results}
\end{table}

\begin{figure}[h]
    \centering
    \includegraphics[width=\columnwidth]{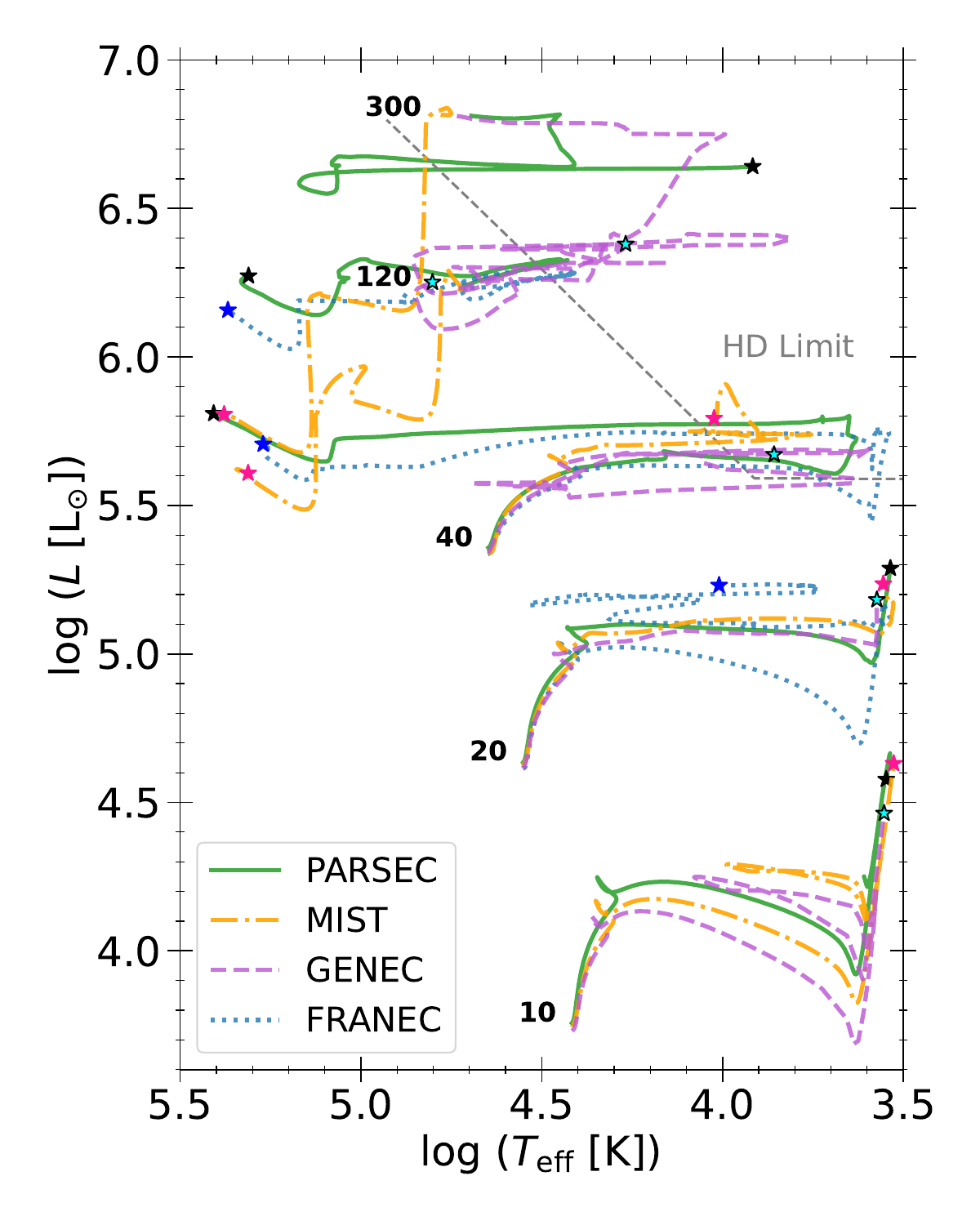}
    \caption{Comparison of non-rotating \textsc{parsec} (green), \textsc{mist} (orange), \textsc{genec} (purple), and \textsc{franec} (blue) tracks in the HR diagram, with \Mz\ = 10, 20, 40, 120, and 300 \msun. Tracks from different codes are computed with the metallicity and input physics parameters listed in Table \ref{tab:code_comparison}. The dashed gray line shows the HD limit. Black, pink, cyan, and blue stars indicate the last model for the \textsc{parsec}, \textsc{mist}, \textsc{genec}, and \textsc{franec} models, respectively.  }
    \label{fig:HRD_Comparison}
\end{figure}

Fig.~\ref{fig:HRD_Comparison} shows the comparison of \textsc{parsec}, \textsc{mist}, \textsc{genec}, and \textsc{franec} tracks for a few selected masses in the HR diagram. 
All tracks begin the MS at about the same \teff\ and luminosity.
The small differences are likely due to the slightly different element partitions and solar-calibrated metallicity.
Generally, there is good agreement in the MS phase among all models up to 40~\msun. 
The different positions in the diagram (particularly the luminosity) at the end of the MS are due to the different overshooting efficiency adopted by different codes (as listed in Tab.~\ref{tab:code_comparison}).
This difference has been found even at lower metallicities between \textsc{genec} and \textsc{parsec} tracks \citep{Sibony2024}. 

In the post-MS evolution, differences increase with \Mz\ since mass loss becomes more efficient.
All 10 \msun\ tracks finish the evolution on the red side of the diagram and with similar final masses.
An analogous result is obtained for the 20 \msun\ models, which agree well with all except for the \textsc{franec} track that performs a blue-loop during the CHeB phase, and finally ends as a YSG, while all other models end as RSGs with $\log \teff \sim 3.5$.
Despite the different paths, the final mass agrees between the \textsc{genec} and \textsc{franec} models, with 9.7 \msun\ and 7.5 \msun, respectively, 
whereas the \textsc{parsec} and \textsc{mist} have larger final masses, 18.3 \msun\ and 14.1 \msun, respectively.

Models with 40 \msun\ begin to disagree more significantly in their post-MS evolution. 
After MS, both the \textsc{franec} and \textsc{parsec} models return to the blue and begin core C-burning before ending evolution as BSGs at $\log \teff \sim 5.4$. 
Conversely, the \textsc{mist} model starts to move to the blue during CHeB, but ends its evolution at $\log \teff \sim 4.1$. 
The \textsc{genec} model crosses the HR diagram twice before completing its evolution close to the \textsc{mist} model. 
At this initial mass, all models except \textsc{mist} become WR stars\footnote{Here, we assume that they become WR when the surface H abundance is below 0.3, in mass fraction.}. 

For the 120 and 300 \msun\ stars, \textsc{parsec}, \textsc{franec}, and \textsc{genec} models move rightward at the beginning of the MS, whereas the \textsc{mist} model has a dramatic decrease in luminosity and increase in \teff. 
These differences can be attributed to different mass loss prescriptions in each model. 
Indeed, \textsc{mist} tracks have the strongest mass loss rates of all the massive star (\Mz\ > 100~\msun) tracks shown and produce the smallest final masses.
Nonetheless, all models become WRs during the MS and move to lower luminosity and higher \teff.
After the MS, the \textsc{parsec} and \textsc{franec} 120~\msun\ tracks evolve very similarly, with the difference that the \textsc{franec} model loses more mass during the evolution.
The evolution of the \textsc{genec} track is a bit different; it evolves toward the red until the ignition of the CHeB finally moves it to the blue, where it dies at $\log\teff \sim 4.8$.
In the post-MS phase, the 300~\msun\ \textsc{parsec} and \textsc{genec} tracks evolve back to lower temperatures after igniting carbon in their core, and finally end as a YSG and a BSG, respectively.
While the 300~\msun\ \textsc{mist} track evolves similarly to the 120~\msun\ and finishes its evolution with $\sim15~\msun$.

The plot shows that stars with $\Mz \geq 40~\msun$ spend some time within the HD limit \citep{Humphreys1979}.
This region of the plot is usually interpreted as the location where stars exceed the Eddington limit and start to lose mass copiously \citep{Grafener2008}.
Indeed, mass loss is a key ingredient that impacts how long these stars stay within the HD limit, and the differences shown in the HR diagram depend mainly on the various adopted mass loss prescriptions.
However, all the models, computed with these codes, with 40 and 120 \msun\ spend less than 0.3 and 0.7 Myr, respectively, beyond the HD limit; therefore, they are very unlikely to be observed\footnote{They are even more difficult to observe if we consider a standard initial mass function as the \citet{Salpeter1955} one.}.
More massive models, with 300~\msun, spend most of their lifetime within the HD limit.
This comparison with the HD limit may suggest more efficient mass-loss prescriptions.
However, it is also worth remembering that we are comparing non-rotating models, and it is known that rotation can strongly impact the evolution of massive stars, the mass lost, and their path in the HR diagram during their life \citep[e.g.,][]{Yusof2013, Limongi2018, Martinet2023}.

Despite the different mass loss rates and the final total mass, the core masses of the \textsc{mist} and \textsc{parsec} models agree very well up to 120 \msun, both having core masses between 0.17 and 0.39 times the initial stellar mass, with the mass percentage of the core increasing with larger \msun.
\textsc{genec}'s core masses are consistently smaller than both \textsc{mist} and Parsec, and is generally around $\sim$ 0.2 times the initial mass of the star. For the 300 \msun\ star, all models have a final mass about equal to the He core.
For the 120 \msun\ models, \textsc{parsec}, \textsc{genec}, and \textsc{franec} match in final masses quite well (32.9 \msun, 30.9, \msun, and 27.8 \msun, respectively), whereas the \textsc{mist} model results in a 11.8 \msun\ final mass. 
This trend is the same in the 300 \msun\ models\footnote{With the exception of \textsc{franec}, which does not have a 300 \msun\ model.} and would suggest different final fates between \textsc{mist} and the other models. 

\subsection{Comparison with observations in the Large Magellanic Cloud}
\label{subsec:comp_w_obs}

The Tarantula Nebula (also known as 30 Doradus) in the LMC has proven to be a treasure trove for hosting massive stars. 
Particularly, the central star cluster R136 hosts some of the most massive single stars ever observed, with the most massive stars exceeding 200 \msun\ \citep{Crowther2010, Bestenlehner2020, Brands2022, shenar2023}.
The Very Large Telescope - Fibre Large Array Multi Element Spectrograph (VLT-FLAMES) Survey \citep{Evans2011} has been used to study this cluster rich in massive stars. 
Multi-epoch optical spectra of over 800 massive stars in the 30 Doradus region have been obtained with this survey. 
Later, \citet{Schneider2018} processed data from the VLT-FLAMES Tarantula Survey to study the formation history and initial mass function of over 200 massive stars in the 30 Doradus Nebula, deriving stellar parameters for hundreds of stars.  
Difficulties due to crowding at the center of this cluster have caused previous surveys to avoid the stars in the center, likely the most massive ones. 
Thanks to the high spatial resolution of the Hubble Space Telescope (HST), \citet{Crowther2016} and later  \citet{Bestenlehner2020} 
studied the physical properties of 55 members of the core of the R136 star cluster. 
Such data have recently been analyzed by \citet{Brands2022} to obtain stellar parameters such as surface abundances and mass loss rates of 53 O-type and 3 WNh-type stars in the center of the cluster. 

\begin{figure}
    \centering
    \includegraphics[width=\columnwidth]{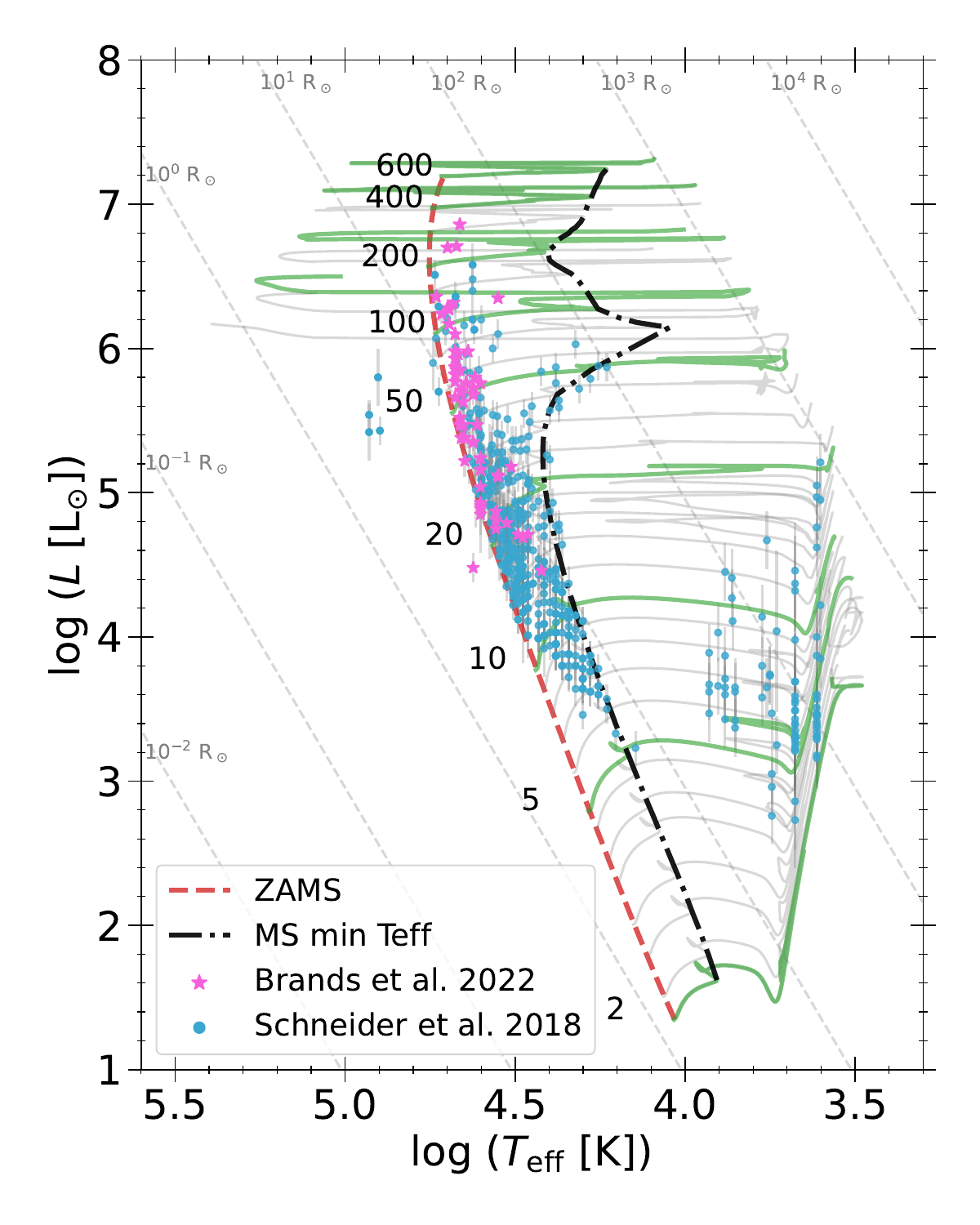}
    \caption{Comparison between data of stars in the Tarantula Nebula and \textsc{parsec} stellar tracks in the HR diagram. 
    Pink stars and blue dots are from  \cite{Brands2022} and \cite{Schneider2018}, respectively.
    Over plotted is the 
    set of tracks with $Z = 0.006$. 
    The dashed red and dash-dotted black lines indicate the ZAMS and TAMS, respectively.}
    \label{Tarantula}
\end{figure}

Here, we compare our models with these observations, which include the most complete massive star sample in a young star-forming cluster known to date.
Fig.~\ref{Tarantula} compares massive stars in the R136 star cluster with our $Z = 0.006$ metallicity set in the HR diagram.
The comparison shows an excellent agreement between evolutionary tracks and data since most stars lie in the MS of our models (highlighted by the red and black lines), the region of the diagram where stars spend most of their lifetime. 
In the mass range $20 \leq \Mz/\msun \leq 50$ ($\log L \sim 5.5$), the width of the MS widens on the HR diagram and follows very well the distribution of data. 
This is due to a combination of mass loss from stellar winds, overshooting from convective cores, and opacity near the surface \citep[as was suggested by][]{Bertelli1984}. 
At the higher mass regime ($\Mz > 50~\msun$, hence, at higher luminosities), the TAMS of our models becomes too cold with respect to data.
These stars have been found to be transitioning between the O spectral type to the WR one, and they occupy a very narrow region of the HR diagram at $\log \teff \sim 4.65$, 
suggesting that these stars should possibly lose more mass during their evolution than that predicted by models \citep[][]{Sabhahit2022}. 
Moreover, all the models have a post-MS evolution that runs at constant luminosity toward even colder temperatures, becoming RSGs, before undergoing blue loops.
As was already mentioned in Sec.~\ref{subsec:comp_w_other_codes}, rotation can be an important ingredient in the evolution of these stars, possibly leading to mass loss enhancement during the MS, causing such very massive stars to become WRs very early in their evolution; finally, narrowing the MS span of these stars in the HR diagram.
On the other hand, recent studies by \citet{Martinet2023} find that in both the rotating and non-rotating cases, the comparison with R136 star cluster data shows a mass loss underestimation and overestimation by models.
Still, there should be a fine-tuning between the various processes at play on top of the observation difficulty to determine the real properties of the stellar surface since these stars experience thick wind mass loss.
Thick winds can occult the star, meaning we are observing this thick, colder layer of ejected material and not the star's surface, which can be much hotter. 
Further investigations about the role of mass loss in very massive stars will be a matter of future work (Shepherd et al. in prep.).

Finally, the fates of the massive stars (130 $\leq$ \Mz/\msun $\leq$ 140) of this set result in PISN explosions. 
Stars of this mass are intrinsically rare at this metallicity, and as such, observations of these events are even less likely.
Even so, current and future telescopes and surveys will be able to aid in detecting and classifying these rare but grand events. 
Particularly, the upcoming Legacy Survey of Space and Time (LSST), which is a 10-year survey that has the capabilities to detect hundreds to thousands of PPISN events, depending on their formation rate \citep[see Chapter 11 of][]{LSSTsciencecollaboration2009}. 

Interestingly, another group of stars lies on the cold side (log \teff\ < 3.8) of the HR diagram, at $2.5 \leq \log L \leq 5$.
This group is composed of stars with $\Mz \leq 20~\msun$, mostly being intermediate-mass stars.
From the evolutionary point of view, after MS, intermediate-mass stars spend most of their time in the CHeB phase as red giants or in the blue loops.
From the comparison, we see that the extension (in \teff) of the blue loops of our tracks is shorter than the width shown by the data. 

Since previous studies have shown that larger envelope overshooting extends the blue loop to hotter effective temperatures \citep{Tang2014, Fu2018}, this comparison may suggest a requirement for an increasing envelope overshooting as a function of mass.
Further investigation on this topic will be reserved for future work.

\subsection{Comparison with observed stellar mass black holes}
\label{subsec:Disc_BH}

In recent years, direct and indirect observations have revealed a continuously growing catalog of BHs in our Universe.
It is worth noting that many of them are the stellar mass BHs (up to $M_\mathrm{BH} \sim 100~\msun$) detected by the LIGO-VIRGO-KAGRA (LVK) collaboration, which provided a plethora of compact object mergers through gravitational waves (GW) detections \citep[][]{LIGO2023}. 
These detections have furnished valuable insights into the properties and origins of these systems, also providing new hints to constrain better properties of massive stars (such as stellar winds).

One of the most noteworthy discoveries from LVK is the BBH merger GW190521. 
With a primary BH mass of 85 $^{+21}_{-14}$ \msun, it is the most massive observed GW progenitor to date \citep{AbbotGW190521.1, Abbot2020GW190521.2}.
This massive BH challenged the existence of the PI mass gap since it lies in the forbidden mass range that goes from about 60 and 130 \msun, as predicted by standard stellar models \citep{Woosley2017}.
Consequently, many possible formation scenarios have been investigated \citep[extensively described in the review by][]{Spera2022}.
Interestingly, several studies have shown that the PI mass gap edges may strongly depend on the uncertain evolution of massive stars \citep{Farmer2019, Farmer2020, Costa2021, Farrell2021, Vink2021, Woosley2021, Volpato2023, Volpato2024}.

Using the new standard set of stellar evolution models presented in this paper, we find that the primary mass of GW190521 can be explained with several models with a metallicity $Z \leq 0.001$.
As is shown in Table~\ref{tab:mgap}, we can form a BH with a maximum mass below the presumed PI mass gap, with $M_\mathrm{BH}$ = 76.5 \msun\ at $Z = 0.001$, consistent within the uncertainty with the observed mass of the primary BH of GW190521.
Since the maximum mass of the lower edge of the PI mass gap increases for lower metallicities, we can also find matching models at lower Z.
It is interesting to note that all the tracks that define the lower edge of the PI mass gap with these metallicities (down to $Z = 10^{-11}$) undergo dredge-up episodes that reduce the He core size, stabilizing them against PI, and finally taking these stars to end their evolution as a FSN.
This makes evident (once again) the important role of convection treatment in massive stars and, in this case, the impact of dredge-up (see Appendix \ref{app:DUP} for more details).
 
Another interesting comparison can be made with recent observations of stellar mass BHs in the Milky Way, namely Cygnus X-1 \citep{Miller-Jones2021}, the first discovered BH, and \textit{Gaia} BH3 \citep{Gaia2024}, the most massive BH discovered (so far) in the Galaxy.

The first, Cygnus X-1, is an X-ray binary with a BH mass of $\simeq 21.2 \pm 2.2~\msun$ with a companion O-star of $\sim 41~\msun$.
The metallicity of this system is thought to be about solar (or more), making the Cygnus X-1 BH the most massive one discovered so far at such a high metallicity.
Since Cygnus X-1 is an interacting system, its $\sim21~\msun$ BH sets a lower limit to the maximum BH mass at solar metallicity.

In our tracks with a correspondent metallicity (i.e., $0.014 \leq Z \leq 0.02$), the most massive BHs formed go from 47 to 34 \msun\ (see Table~\ref{tab:mgap}),
still higher than that of Cyg X-1 BH but agrees with other single stellar models from various authors in the literature.
For instance, our models with $Z = 0.014$ predict a maximum BH mass that agrees with the 48 \msun\ (for non-rotating stars) found with \textsc{genec} by \citet{Martinet2023}, but is larger with respect to that obtained by \citet{Bavera2023}, \citet{Romagnolo2024} and \citet{Vink2024}, who found 35, 28, and 30 \msun, respectively.
Such results instead agree with the maximum BH mass of our tracks with $Z = 0.02$, suggesting that these authors assume more efficient stellar winds prescription than ours. 
Finally, we note that the final mass spectrum for stars with $Z = 0.014$ has a trend of increasing final BH mass with increasing \Mz\ (up to 270 \msun), consistent with the one found by \citet{Romagnolo2024}, but not with others \citep{Bavera2023, Martinet2023, Vink2024}.

\textit{Gaia} BH3 is a 33 \msun\ BH that has recently been discovered in our Galaxy, making it the most massive BH of stellar origin discovered so far via electromagnetic counterparts \citep{Gaia2024}. 
The discovery was possible thanks to it being in an 11.6-year binary with a 0.76 \msun\ giant with low metallicity of [M/H]=-2.21.
There are already several studies to explain the formation of \textit{Gaia} BH3, some suggesting that it could be formed via dynamical interactions \citep{El-Badry2024, Pina2024}, others which favor the isolated non-interacting evolution \citep{Iorio2024}.
If the BH progenitor star shared a similar metallicity as its companion (i.e., $Z \sim 0.0001$), our models suggest that \textit{Gaia} BH3 could have formed with an initial mass of 35 \msun. 
This star produces a BH of 34 \msun\ and maintains its radius smaller than 700 R$_{\odot}$, well below the present minimum separation of the binary, which is 1000 R$_{\odot}$, and the minimum radius needed for such a binary to interact via mass transfer episodes \citep[$\sim 800$ R$_{\odot}$,][]{El-Badry2024}. 
Therefore, with our stellar models, such a system could be formed through a non-interacting isolated binary, confirming the results by \citet{Iorio2024} using \textsc{mist} tracks \citep{Choi2016}.

Also, higher mass models ($\Mz \sim 110-120$ \msun) can form a BH of about 30 \msun. However, in that case stars become RSGs, and a common envelope interaction would be very unlikely to be avoided.

\section{Conclusions}
\label{sec:conclusion}

In this paper, we have presented our new standard library of non-rotating stellar evolutionary tracks computed with \textsc{parsec}, as well as several additional results this collection of models delivers us.
The sets of tracks are composed of over $\sim 1100$ models ($\sim 2100$ including pure-He tracks) with 13 different metallicities ranging from $Z = 10^{-11}$ to $Z = 0.03$, spanning Pop I, Pop II, and Pop III, and considering initial masses from 2 \msun\ to 600 (2000) \msun.
We summarize the main results of this study below:
\begin{itemize}

  \item There are several differences in the input physics with the last release of \textsc{parsec} stellar models of massive stars \citep{Chen2015}. 
  The main ones regard the new mixing scheme included in \textsc{parsec v2.0} \citep{Costa2019a, Nguyen2022}, the updated EOS including pair creation, and the updated stellar winds for WR stars. 
  All these updates allow us to follow the evolution of such stars more realistically until the most advanced pre-SN phases.
  A detailed comparison with previous models with the same metallicity is given in Appendix~\ref{app:Comp_wParsecv1.2s}.
  
  \item From the comparison of Pop III, Pop II, and Pop I stars in the HR diagram, we confirm that stellar winds and the convection treatment are the main drivers of the evolution of non-rotating massive and very massive stars.
  In agreement with other studies, we find that the ZAMS and the MS are hotter as the metallicity decreases.
  Metal-poor and metal-free stars with \Mz\ $\leq 200$ \msun\ ignite the He shortly after the end of the MS, at variance with more metallic stars that ignite after first becoming RSGs.
  Interestingly, we find that stars with metallicity below $Z = 0.001$ (namely, not affected by winds) and \Mz\ $\geq 200$ \msun, end their MS as RSG stars.
  Other authors have also found this result \citep[such as][]{Tanikawa2022}, but the actual transition mass depends strongly on the stellar physics adopted.

  \item We computed the final fate of stars and build the so-called mass spectrum (Fig.\ref{fig:Mass_gaps_pops}), finding (as expected) that the PI mass gap changes for different initial metallicities (Table~\ref{tab:mgap}).
  This is due to the various processes acting during the evolution of massive stars, which lead stars with the same \Mz\ to explode differently, depending on the metallicity.
  Moreover, we find that the transition between different final fates is not monotonic and that the combined BHs PI mass gap goes from $\sim$100 to $\sim$130 \msun.

  \item We present the computed stellar wind and explosive ejecta, showing how much the two different types contribute to the total ejecta for varying initial masses and metallicities. 
  Again, we find that stellar winds play an increasing role as $Z$ increases. 
  Still, we also find that at solar-like metallicity, the contribution of PISN ejecta remains important, particularly for the processed elements up to Fe.
  
  \item By analyzing the ionizing radiation from our models, we find a general good agreement with the result from other works.
  From the comparison of the integrated photon numbers, we find that very metal-poor stars ($Z = 10^{-6}$) can emit a similar (or even higher) amount of ionizing photons with respect to Pop III stars, suggesting that the radiation feedback to the environment during their formation and evolution, could be similar.

  \item The comparison of massive stellar tracks computed with other codes shows a generally good agreement.
  The various assumed treatments for convection and stellar wind prescriptions are the main cause of the differences that, in turn, increase with the initial mass.
  This confirms the uncertainties regarding the evolution of massive stars, underscoring the need for continued investigation into the topic.
  
  \item We use the LMC Tarantula Nebula massive stars data collection to test our models.
  We find a good agreement between the models and the data MS's width in the HR diagram.
  We also notice that at higher masses (\Mz > 50 \msun), the models become too cold with respect to data, probably requiring stronger winds or high rotation rates to shrink the MS width.  
  Moreover, the small blue loops of intermediate-mass stars models with respect to the data position in the diagram suggest the need for a higher envelope overshooting parameter than that used for our tracks.
  These aspects and others will be matters of further study in follow-up investigations.

  \item By comparing our results with observed stellar mass BH, we find that we can well reproduce properties of observed BHs, such as Cygnus X-1 and \textit{Gaia} BH3 primaries with our stellar models. 
  Moreover, we find that the maximum mass at solar and galactic metallicity is about $40\pm6$ \msun, in agreement with other results in the literature.
\end{itemize}

In the appendices, we provide more plots to show stellar properties, give details about strong dredge-up episodes in massive stars, compare the new tracks with older computed with previous \parsec\ versions, present our grids of pure-He stars, and summarize important features for a subset of metallicities and masses in Table \ref{tab:stellar_properties}.

\section{Data availability}
All results, including stellar tracks, tables of final fates and stellar ejecta, and ionizing photons, are released in the public \textsc{parsec} web database at \url{http://stev.oapd.inaf.it/PARSEC/}.

\begin{acknowledgements}
The authors wish to express their deepest gratitude to Professor Paola Marigo, who passed away on October 20, 2024. Her memory will be cherished forever, and her invaluable contributions, passion, and humanity will continue to inspire us and all those fortunate enough to have the privilege of working with her.
We thank the referee for his useful comments.
The authors acknowledge Iorio G., Mapelli M., and Spera M. for the helpful discussions. We thank Sibony Y., Yusof N., and Ekstr{\"o}m S. for providing data of \textsc{genec} tracks.
GC acknowledges financial support from the Agence Nationale de la Recherche grant POPSYCLE number ANR-19-CE31-0022.
GC acknowledges support from the European Research Council (ERC) through the ERC Consolidator Grant DEMOBLACK, under contract No.~770017.
We acknowledge partial support from the ERC Consolidator
Grant funding scheme, project STARKEY, G.A. No. 615604.
AB and PM acknowledge the Italian Ministerial grant PRIN2022, “Radiative opacities for astrophysical applications,” no. 2022NEXMP8.
MT and GP acknowledge financial support from the Supporting TAlent in ReSearch@University of Padova (STARS@UNIPD) for the project ``CONVERGENCE: CONstraining the Variability of Evolved Red Giants for ENhancing the Comprehension of Exoplanets''. LG and SZ acknowledge support from an INAF Theory Grant 2022.
X.F. thanks the support of the National Natural Science Foundation of China (NSFC) No. 12203100 and the China Manned Space Project with No. CMS-CSST-2021-A08.
AM acknowledges financial support from Padova University, Department of Physics and Astronomy Research Project 2021 (PRD 2021) and from Bologna University, ``MUR FARE Grant Duets CUP J33C21000410001''.
Y.C. thanks the support of the National Natural Science Foundation of China (NSFC) No. 12003001, Natural Science Research Project of Anhui Educational Committee No. 2024AH050049 the Anhui Project (Z010118169).
\end{acknowledgements}

\bibliographystyle{aa} % style aa.bst
%\bibliography{biblio} % your references Yourfile.bib


\appendix

\section{Pre-Main Sequence}
\label{app:PMS}

\begin{figure}[h]
    \centering
    \includegraphics[width=0.87\columnwidth]{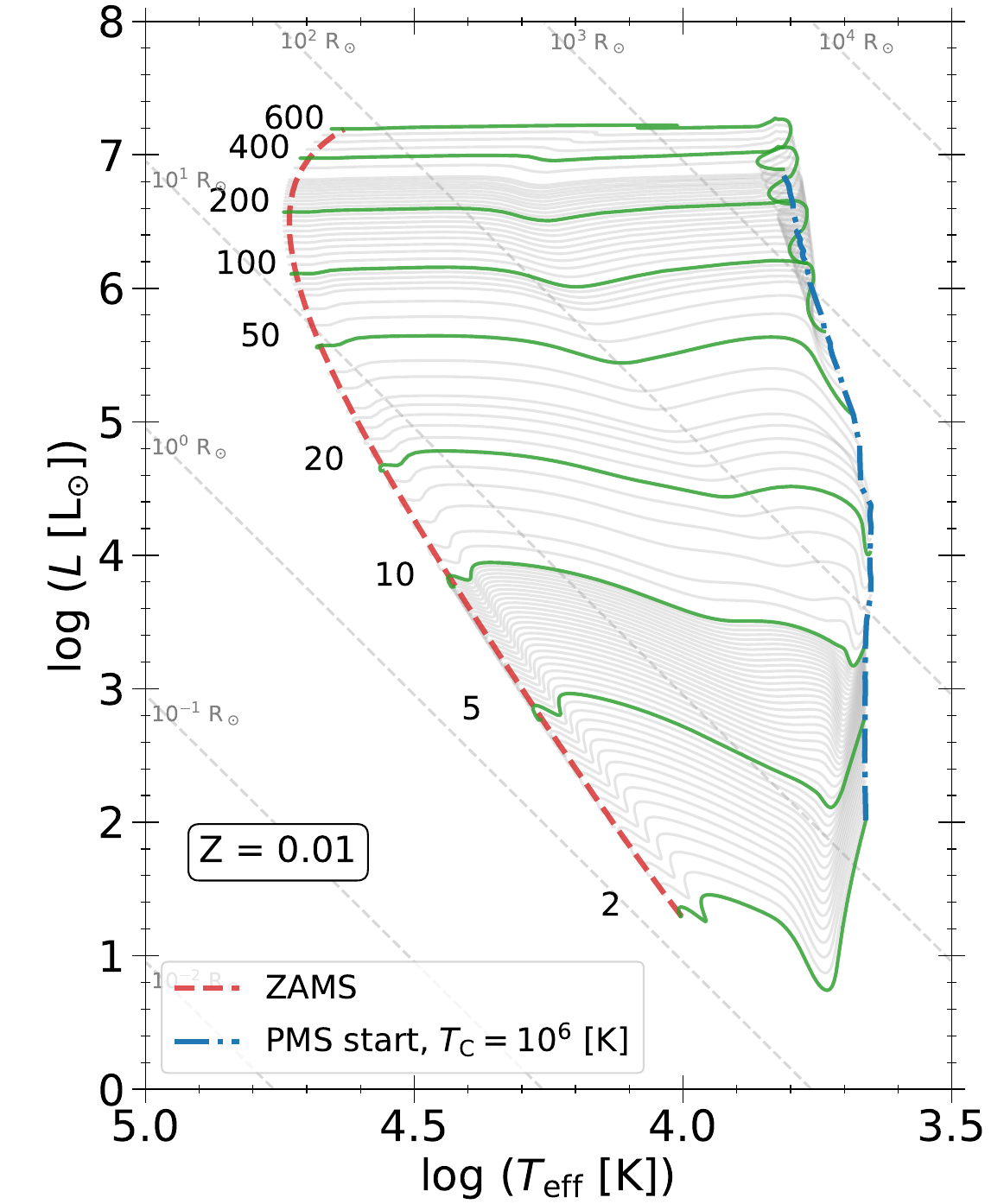}\\
    
    \caption{Pre-Main Sequence evolution of stars with $Z=0.01$. 
    The dashed red line indicates the ZAMS, while the dash-dotted blue line indicates the PMS start position in the diagram.
    Selected tracks are shown in green, and the corresponding initial mass is indicated near their ZAMS.
    Diagonal dashed gray lines show constant stellar radii as labeled.}
    \label{fig:app_PMS}
\end{figure}

To build the PMS of our tracks, we follow the contraction from an expanded model near the Hayashi line.
The evolutionary tracks begin with a fully convective hydrostatic model with a central temperature of $T_\mathrm{C} = 10^6$ K, and evolve at constant mass.
Fig.~\ref{fig:app_PMS} shows an example of evolution in the HR diagram of our tracks with $Z=0.01$.
The PMS evolution changes with \Mz\ because of the different structure response to the ignition to deuterium first and lithium later.
In all cases, stars contract, reaching the ZAMS, where the star's gravity is balanced by the energy released by the hydrogen burning.
We devolve the investigation on the accretion process during the PMS phase to future work \citep[although some work has already been done for low-mass stars, e.g.,][]{Fu2015}.

\section{Stellar properties}
\label{app:Stellar_prop}

\begin{figure}[h]
    \centering
    \includegraphics[width=0.87\columnwidth]{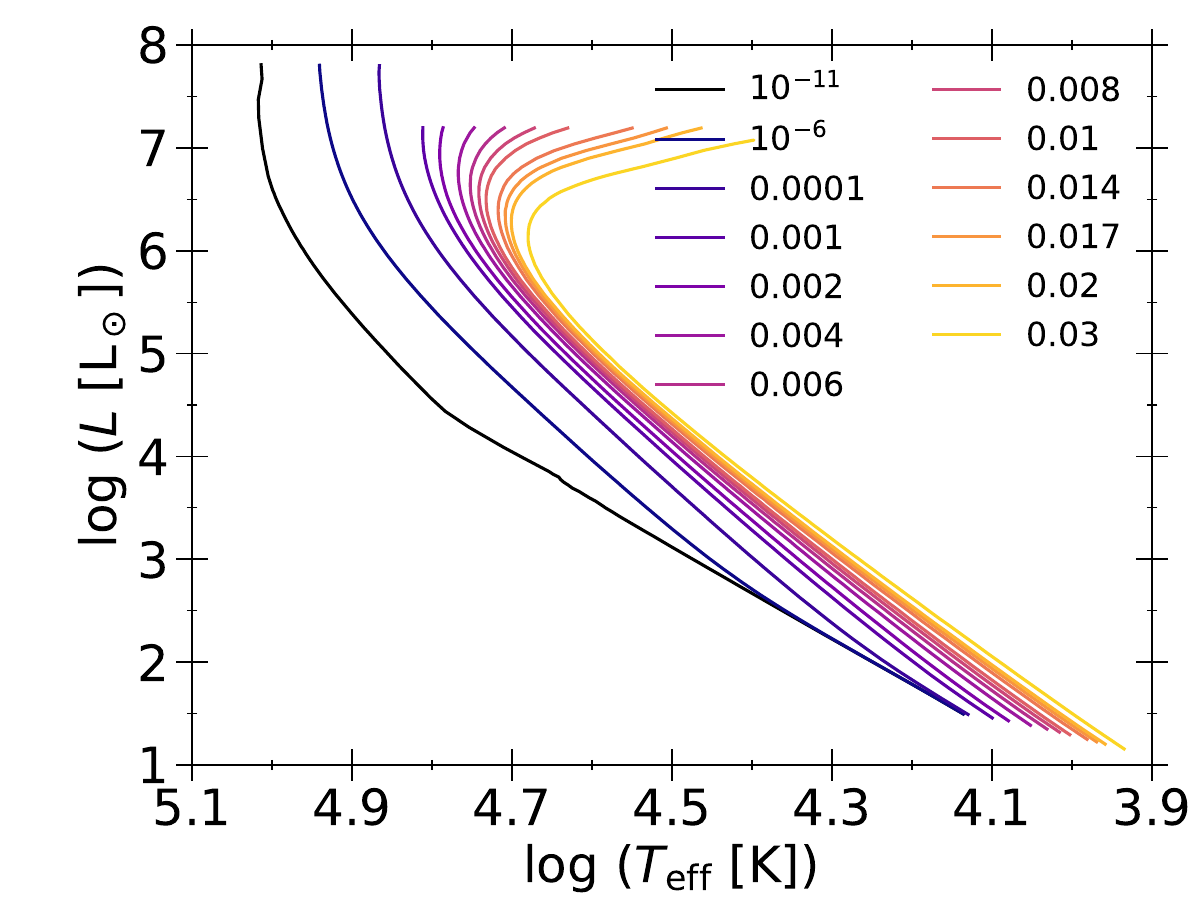}\\
    \includegraphics[width=0.87\columnwidth]{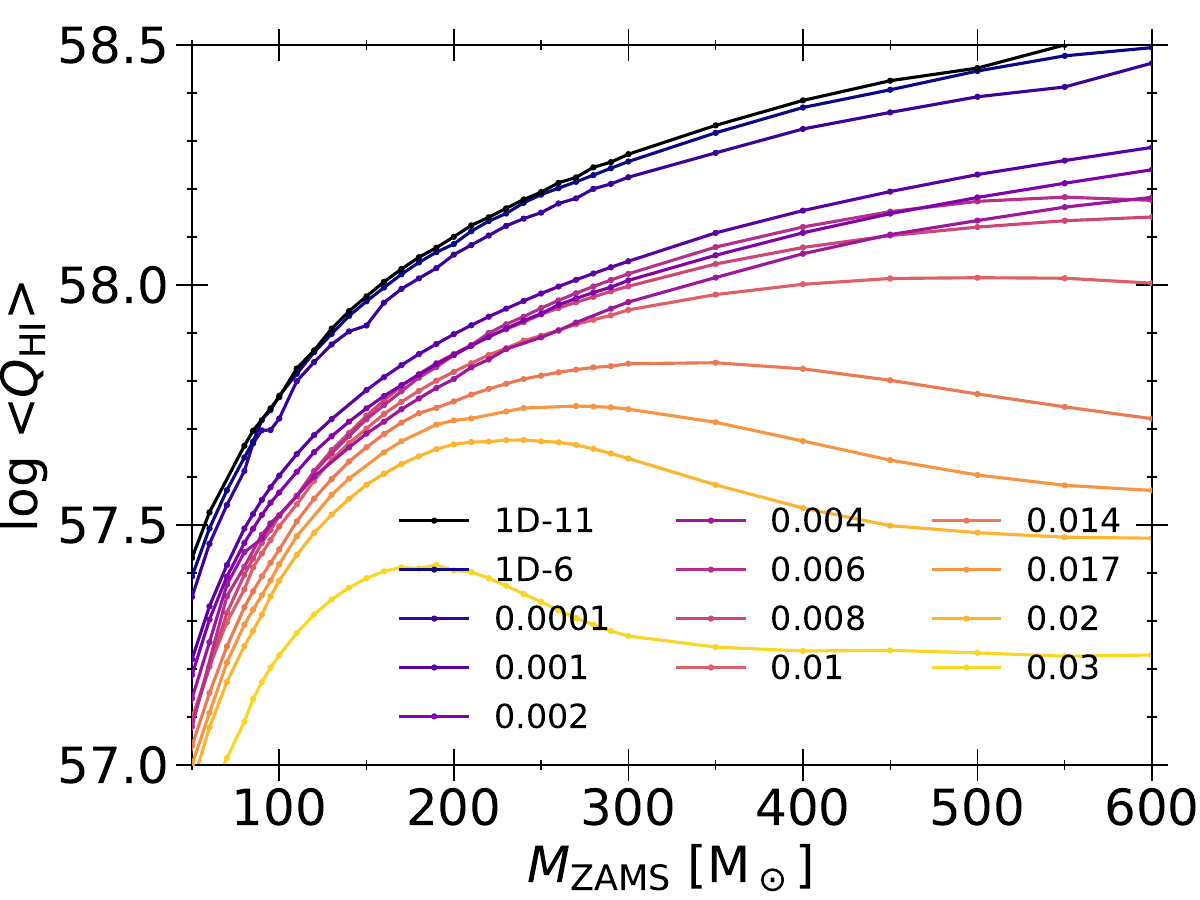}
    
    \caption{Top: Comparison of ZAMS at different metallicity, indicated by the color and line style in the HR diagram.
    Bottom: The average rate of ionizing photons over the lifetimes versus the initial mass for different metallicities.}
    \label{fig:app_ZAMS}
\end{figure}

\begin{figure*}[t]
    \centering
    \includegraphics[width=0.87\textwidth]{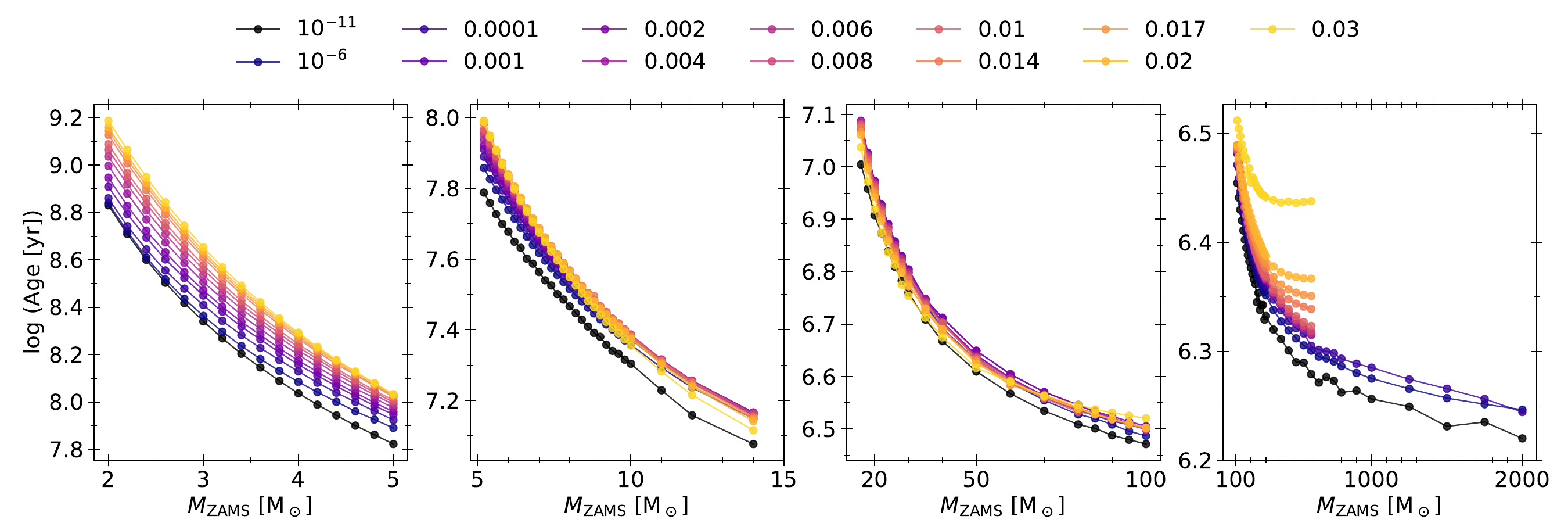}
    \caption{Comparison of stellar lifetimes at different metallicity, indicated by the color.}
    \label{fig:app_lifetimes}
\end{figure*}

Here, we show some properties of the grids in our database.
The upper panel of
Fig.~\ref{fig:app_ZAMS} shows the ZAMS position in the HR diagram of stars with different initial metallicity. 
It clearly shows that the ZAMS position becomes hotter as the metallicity decreases.

The lower panel shows the average number of photons emitted per second over the entire lifetime for stars with different initial metallicities.
This comparison shows that the amount of ionizing photons emitted decreases as the metallicity increases.
It also shows that very-metal poor Pop II stars can emit a similar amount of ionizing photons as Pop III stars.

Fig. \ref{fig:app_lifetimes} shows the total lifetimes of stars at different mass regimes and metallicity.
The comparison shows a monotonic increase in the lifetimes with metallicity for stars in the mass ranges with $\Mz \lesssim 8~\msun$ and $\Mz \gtrsim 100~\msun$.
In the middle mass range, the lifetime trend is not monotonic with $Z$. 
This is mainly due to opacity and mass loss dependence on $Z$ for the intermediate-mass and massive regimes, respectively.

\section{Deep dredge-up events and core-envelope interactions in advanced phases} 
\label{app:DUP}

\begin{figure*}[h!]
    \centering
    \includegraphics[width=0.87\textwidth]{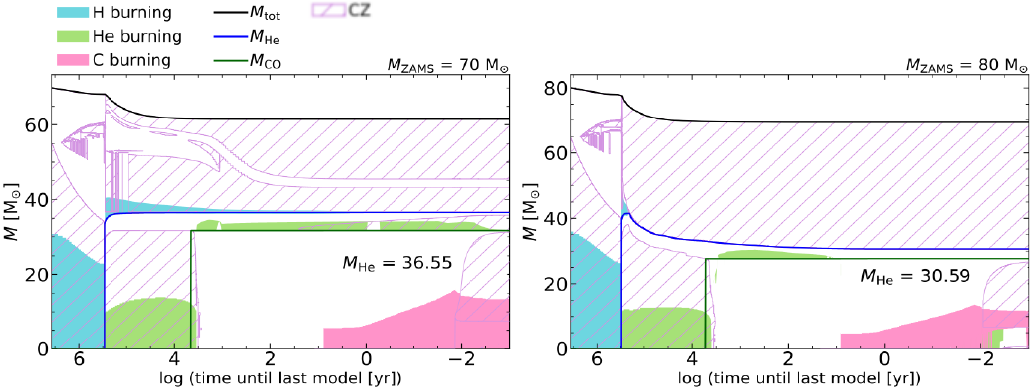}
    \caption{Kippenhahn diagrams for $\Mz = 70~\msun$ (left) and $\Mz = 80~\msun$ (right) with $Z = 0.001$. 
    The Convective zones are indicated by purple hashed lines, and vertical purple lines indicate adjacent convective regions. Burning zones of H, He, and C are indicated by solid-shaded blue, green, and pink regions, respectively. The He core and CO core masses are represented by blue and green lines, respectively.}
    \label{fig:app_kipp_Z0.001}
\end{figure*}

The dredge-up is an opacity-driven mechanism that occurs when the star's envelope expands and cools, causing the opacity to rise. 
The envelope's temperature gradient becomes superadiabatic, and the base of the convective envelope extends inward into the inner layers.

At low-metallicity, $Z \leq 0.001$, massive stars can experience dredge-up processes when they become RSGs, in some cases becoming particularly deep during the CHeB phase.
In these cases, the bottom of the convective envelope can cross the H-He discontinuity, penetrating the He core. 
When this happens, the dredge-up can lead to a significant decrease in the He core mass, making these models more stable against the PI, and impacting the final fate.
This aspect and the role of the envelope overshooting in the dredge-up intensity has already been discussed by \citet{Costa2021}.

Here, to convey how this process affects the star's evolution and impacts the final fates, in Fig. \ref{fig:app_kipp_Z0.001}, we show the Kippenhahn diagram of two models with $Z = 0.001$ and different initial mass, 70 and 80 \msun, which does not and does experience strong dredge-up, respectively.

The left panel of Fig.~\ref{fig:app_kipp_Z0.001} shows how, after the He ignition, the \Mz\ = 70~\msun\ star develops large intermediate convective regions.
Although the deeper region seems close to the H-He discontinuity, it is always detached from the core border, being fueled by a thin H-burning layer above it.
In this case, convection does not penetrate the He core, and the star ends its evolution with $\MHe \sim 36.5~\msun$, thus entering the PI regime.
On the other hand, the right panel shows how, in the \Mz\ = 80~\msun\ star, the convective envelope deepens in the He-core during the CHeB. 
As the star evolves in this phase, the bottom border of the convective envelope penetrates the core and drags material from the core to the envelope.
This material enriches the outer layers and is mixed up to the stellar surface, altering the star's chemical composition. 
The 80~\msun\ star experiences a decrease in the He core mass due to dredge-up and finishes its evolution with $\MHe \sim 30.5$ \msun, finally avoiding PI.
This different evolution explains the island of FSN fates between 80 and 100~\msun\ in Fig.~\ref{fig:Remnant_type_grid}.
It is worth saying that the evolution of massive stars is particularly sensitive to the physics assumed, and the evolutionary behavior can be chaotic, as tiny variations in the initial conditions can lead to big evolutionary differences. 
Within this context, efficient dredge-up can play a significant role in the determination of the stars' final fate and remnant mass. 
For instance, Table~\ref{tab:stellar_properties} shows that our Pop III 100~\msun\ star ends its evolution with a He core mass of $\MHe \sim 33~\msun$, smaller than that of the 100~\msun\ with $Z=10^{-6}$, $\MHe \sim50~\msun$, due to a stronger dredge-up event.
Therefore, the Pop III star ends its evolution as with a FSN, leaving a BH of $\sim 100~\msun$. On the other hand, the $Z = 10^{-6}$ 100 \msun\ star finishes as a PPISN, leaving a BH of 43 \msun.

As was previously stated, dredge-up events also affect nucleosynthesis, for example, by boosting the production of primary nitrogen via the CNO cycle when the H-burning shell is polluted with the products of the He-burning.
This nitrogen can be considered primary for low metallicity stars because the carbon and oxygen used in the CNO cycle were produced by the star and not present at birth, similarly to the primary N enrichment induced by rotation, as is noted in \citet{Tsiatsiou2024}.
The 70~\msun\ star reaches the pre-SN stage with an enriched surface with $X\mathrm{^S_H} \sim 0.49$, $X\mathrm{^S_{He}} = 0.50$, $X\mathrm{^S_N}$ = 0.0006, and $X\mathrm{^S_C} \approx X\mathrm{^S_O} = 10^{-5}$.
While the final surface composition of the 80~\msun\ star is $X\mathrm{^S_H}$  = 0.34, $X\mathrm{^S_{He}}$ = 0.52,  $X\mathrm{^S_C}$ = 0.045,  $X\mathrm{^S_N}$ = 0.05, and  $X\mathrm{^S_O}$ = 0.036, the strong enhancement of CNO elements in the case of the dredge up are evident.

\section{Comparison with PARSECv1.2s}
\label{app:Comp_wParsecv1.2s}

\begin{figure*}[h!]
    \centering
    \includegraphics[width=0.87\textwidth]{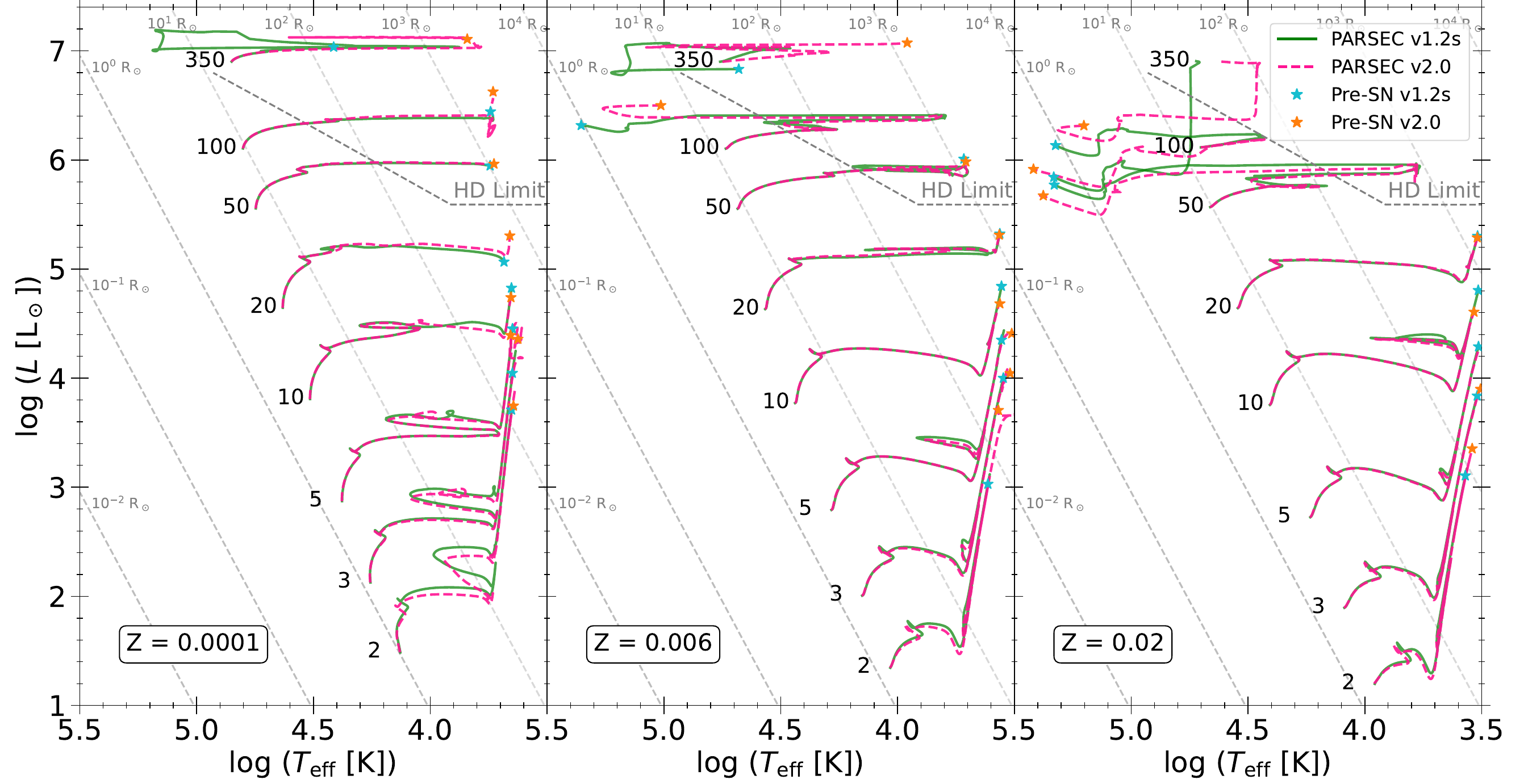}
    \caption{HR diagrams of \parsec\ versions 1.2s (solid green) and v2.0 (dashed pink) presented in this work for three metallicities. 
    Cyan and orange stars indicate the final pre-SN location in the HR diagram of stellar tracks computed with \parsec\ v1.2s and v2.0, respectively. 
    The initial mass of each is indicated near the ZAMS in solar masses.
    Diagonal dashed gray lines indicate constant stellar radii as labeled.
    It is worth noting that for $Z = 0.0001$, the helium abundances are slightly different $Y = 0.249$ and $Y = 0.2487$ for v1.2 and v2.0, respectively.
    }
    \label{fig:Parsec_versions}
\end{figure*}

The new grid of models encompasses a broader spectrum of initial stellar masses with a finer mass spacing than previous releases of \textsc{parsec} stellar tracks. 
Additionally, it extends to a lower metallicity range than previous releases, enhancing the applicability of \parsec\ v2.0 models. 

Several differences in the input physics exist between the two versions of the code, and the new setup has already been presented in Sec.~\ref{sec:Meth}.
The new input physics mainly affects the evolution of massive stars in the advanced phases. 
To illustrate this, Fig. \ref{fig:Parsec_versions} shows the comparison of selected tracks computed with \textsc{parsec} v1.2 and \textsc{parsec} v2.0 in the HR diagram at different metallicities.
From the comparison, we find a general very good agreement for all initial masses and metallicities.
Stars with $\Mz \leq 20~\msun$ evolve in a slightly different way mainly due to the different mixing scheme, which now simultaneously solves the nuclear reaction network and the mixing using an implicit diffusive scheme.
The small differences accumulate along the evolution and tend to have slightly hotter TAMS positions and less luminous blue loops in the CHeB phase for tracks computed with \textsc{parsec} v2.0.
At higher masses ($\Mz \geq 50~\msun$), differences can be bigger because of the highly sensitive response of the convective regions to initially small variations, and because of the different stellar winds prescriptions (in the $Z > 0.0001$ cases).

At $Z = 0.0001$, the 50 and 100~\msun\ stars run almost overlapped in the HR diagram and end their evolution as RSGs, while the 350~\msun\ stars' evolutions diverge.
This is likely due to the slightly different mixing scheme that causes the old track to experience a dredge-up that lowers the surface hydrogen, making the star a WR (with $X\mathrm{^S_H} < 0.3$) with strong stellar wind.
At variance, the new track does not experience such a strong drop of the surface hydrogen, and finally, it experiences less intense stellar winds. 
At the end of the evolution, the \textsc{parsec} v1.2s track is on the blue side of the HR diagram with a mass of $\sim 166~\msun$ and a He core of $160~\msun$.
While the \textsc{parsec} v2.0 track dies as an RSG, after performing a blue loop during the CHeB phase, with a total mass of $\sim 340~\msun$ and a He core of $\sim 205~\msun$.

At higher metallicity, it is more evident that the discrepancies arise due to different mass loss prescriptions for WR stellar winds, as the evolution of all stars with $\Mz \geq 50~\msun$ is almost identical until they become WR and diverge.
For these stars, the locations of TAMS and the start and end of CHeB in the HR diagram are nearly identical.
The final locations change due to the different final masses of the two models of stars.
In particular, we find that all \textsc{parsec} v1.2s tracks end their lives less massive than \textsc{parsec} v2.0 ones, with, in turn, smaller He cores. 

Another difference we find is that massive models from v1.2 spend longer times on the MS, resulting in slightly longer total lifetimes compared to v2.0 ones. 
This is likely due to the different mixing schemes used. 

Finally, \textsc{parsec} v1.2 models evolve massive stars (\Mz\ $\geq$ 10 \msun) up to central carbon ignition. 
In contrast, the new models presented here are evolved up to the advanced phase of either oxygen burning or entering the PI regime, providing insights into the final stages of stellar evolution and aiding in our understanding of supernova and BH progenitors.  

\section{Pure He stars}
\label{app:PureHe}

\begin{figure*}[h]
    \centering
    \includegraphics[width=0.87\textwidth]{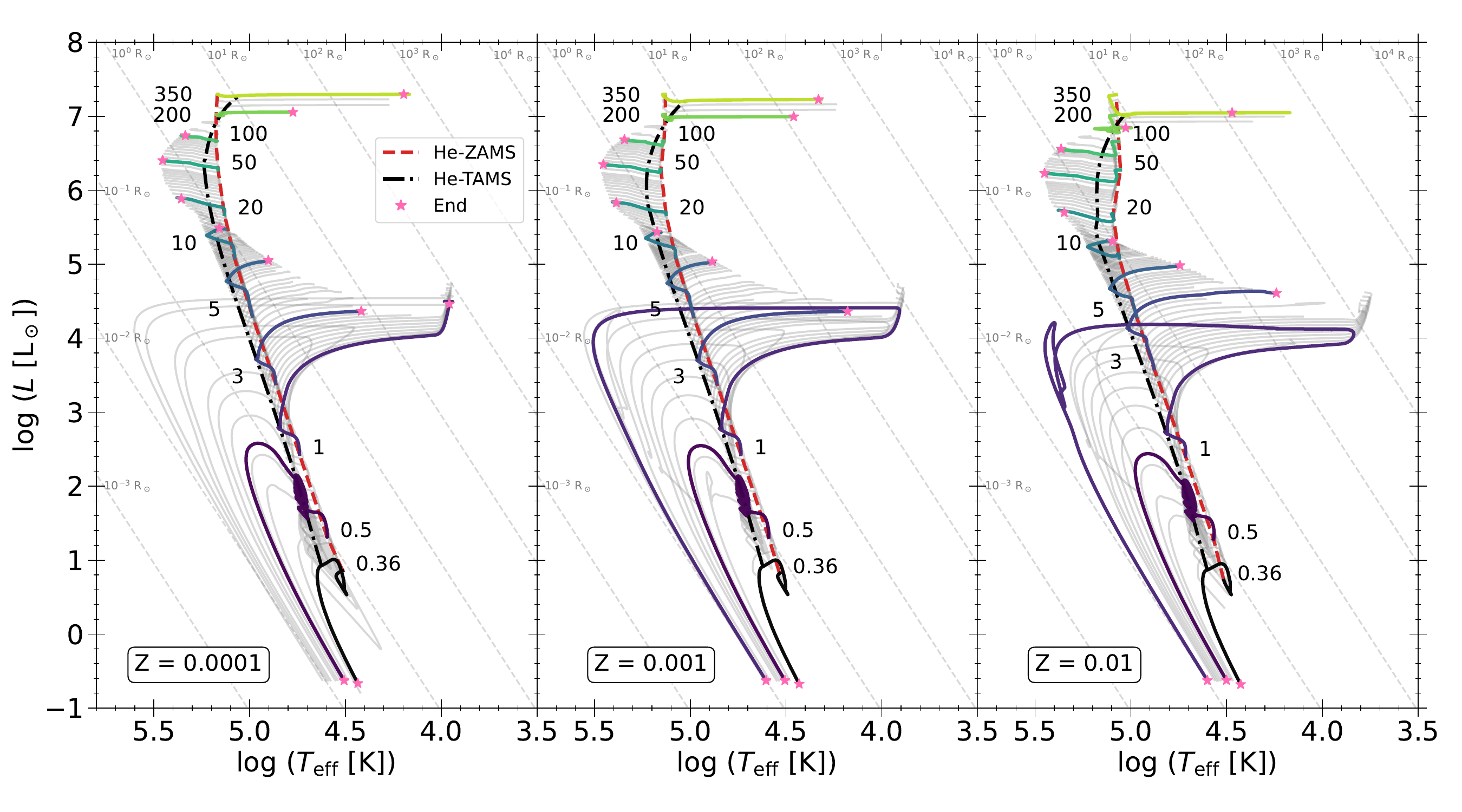}
    \caption{HR diagrams of pure-He tracks of $Z = 0.0001$ metallicity (left), $Z = 0.001$ (center), and $Z = 0.01$ (right). Dashed red and dash-dotted black lines indicate the beginning and end of the He-MS. Pink stars indicate the final location for a few highlighted masses, whose mass is indicated close to the He-ZAMS position in \msun. Diagonal dashed gray lines indicates constant stellar radii as labeled.}
    \label{fig:HRD_pureHe}
\end{figure*}

Here, we present a new grid of pure-He stars computed with \textsc{parsec} v2.0.
These models are not only practical themselves but are also particularly useful for population synthesis codes that include binary evolution processes and use precomputed stellar track tables.
When the stellar envelope is totally stripped by binary interactions, the outcome stripped star is a bare helium core, and using standard single evolutionary tracks to follow the subsequent evolution is not accurate enough. 
Pure-He stars should be used instead, with proper interpolations on the properties of such stars after the stripping \citep[e.g.,][]{Spera2019}. 

The new grid of pure-He tracks includes stars in the mass range of 0.36 $\leq \Mz/\msun \leq$ 350 with mass steps as is indicated in table \ref{tab:pure_He}.
The metallicities of these tracks are $Z = 10^{-6}$, 0.0001, 0.0002, 0.0005, 0.001, 0.002, 0.004, 0.008, 0.01, 0.02, 0.03, and 0.05.
While the initial He is simply given by $Y = 1 - Z$.
The input physics of pure-He tracks is the same as that described in Sec.~\ref{sec:Meth}, with a difference for stellar winds.
For pure-He stars, we use the mass loss prescriptions given by \citet{NugisLamers2000}, including also a dependence on the Eddington ratio \citep{Vink2011}.
Pure-He stars are initialized from stripped standard stars at the beginning of the CHeB phase.
After removing the envelope, the stellar mass is gradually changed to reach the required initial mass at fixed chemistry.
When pure-He stars reach the He main sequence (He-MS), the chemical evolution is restarted.

\begin{table}[]
    \caption{Pure-He Masses}
    \begin{center}
   
    \begin{tabular}{c|c}
        \toprule
        Mass Range & Step Size  \\
         $[\msun]$ & $[\msun]$ \\
        \midrule
        0.36 - 0.5 & 0.02 \\
        0.5 - 1.0 & 0.05 \\
        1.0 - 2.0 & 0.1 \\
        2.0 - 9.0 & 0.2 \\
        9.0 - 12.0 & 0.5 \\
        12.0 - 20.0 & 1.0 \\
        20.0 - 40.0 & 5.0 \\
        40.0 - 100.0 & 20.0 \\
        100.0 - 350.0 & 50.0 \\
        \bottomrule
    \end{tabular}
    \end{center}

    \footnotesize{Column 1: Mass range. Column 2: Corresponding mass step}
    \label{tab:pure_He}
\end{table}

The evolution of these stars is different with respect to `standard' stars; this can be seen in Figure \ref{fig:HRD_pureHe}, which shows the HR diagram of pure-He tracks for three metallicities. 
All tracks with masses $\Mz < 200~\msun$ during the He-MS increase the \teff\ and luminosity until central He exhaustion is reached. 
On the other hand, during the He-MS, stars above $\sim$ 200 \msun\  evolve to cooler \teff\ at constant or decreasing luminosity, depending on the metallicity. 
This is because of the enhanced mass loss due to metallicity-dependent winds. 

After the CHeB phase, stars with $\Mz \leq$ 15 \msun\ evolve back to cooler temperatures.
While stars with 15 < \Mz/\msun\ < 200 continue to evolve toward hotter temperatures.
This behavior determines the final positions of stars in the HR diagram. 

After the He-MS, models with masses below $\sim$ 1.6 \msun\ become strongly degenerate and move to the `AGB' phase, in which they burn He in a shell above the CO core.
The evolution and the outcome of these models depend on the competition between the He-shell burning and the mass loss stripping.
Models with $\Mz \lesssim$ 1 \msun\ are strongly stripped and evolve to the white dwarf (WD) cooling sequence. 
Stars with initial masses below about 1.6 \msun\ lose enough mass from stellar winds making their total mass fall below that necessary for carbon ignition and, therefore, they will end as WDs as well.
Stars with mass $1.6 \leq \Mz/\msun \leq 2.4$ reach milder degenerate conditions in the core and are able to ignite carbon off-center, likely ending as CO-WDs.
Models with \Mz\ from 2.4 \msun\ to 3.0 \msun, reach a final M$_{CO} \simeq 1.37-1.38~\msun$, close to the threshold between becoming an O-Ne-Mg WD and an electron-capture SN \citep{Miyaji1980, Nomoto1984, Takahashi2013}.
This transition initial mass depends on the metallicity, with larger initial masses required for increasing metallicity.

Stars above 3~\msun\ ignite C in non-degenerate conditions and evolve until the oxygen-burning phase.
Pure-He massive stars with masses above 30~\msun\ start to experience an efficient pair-creation process during the oxygen-burning phase.
As for standard stars computations, we stop the evolution of such massive Pure-He stars when the stability criterion given in Eq.~\ref{eq:PI} is no longer satisfied.

The grids of Pure-He stars are also publicly available in the \parsec\ database.

\section{Extra tables}
\label{app:extra_tables}
Table~\ref{tab:stellar_properties} lists the properties of selected tracks at the end of the evolution for different metallicities. 
This table collects selected results from our stellar tracks and ejecta tables database.

\begin{table*}[h!]
\begin{center}
\caption{Properties of selected massive stars for seven metallicities.}
    \begin{tabular}[c]{c c c c c c c c c c}
    \toprule
    
    \Mz\ & $t_\mathrm{lifetime}$ &  $M_\mathrm{Pre-SN}$  & Fate & $M_\mathrm{Remnant}$ & $M_\mathrm{He}$ & $M_\mathrm{CO}$ &  $X\mathrm{^c_{O}}$ & $\log T\mathrm{_c}$ & $\langle \Gamma_1 \rangle_\mathrm{Pre-SN}$ \\
         
        [\msun] & [Myr] & [\msun]  &  &[\msun] & [\msun] & [\msun]   &   & [K] &  \\
        (1) & (2) & (3) & (4) & (5) & (6) & (7) & (8) & (9) & (10)  \\
        \midrule
        \multicolumn{10}{c}{$Z = 10^{-11}$} \\
        \midrule
          14  &     11.93  &    14      &    CCSN    &     1.49  &     3.84  &     2.79  &     0.01  &        9.35  &      1.44  \\
          20  &     8.08  &    20      &    CCSN    &     1.57  &     6.73  &     4.97  &     0.00     &        9.40  &      1.42  \\
          30  &     5.75  &    30      &    FSN     &    29.50    &    12.08   &     9.79  &     0.05  &        9.39  &      1.41  \\
          40  &     4.64  &    39.99  &    FSN     &    39.50    &    16.91    &    15.08   &     0.02  &        9.45  &      1.38  \\
          50  &     4.07  &    49.99  &    FSN     &    49.50    &    22.43   &    19.65   &     0.02  &        9.45  &      1.36  \\
          80  &     3.22  &    79.99  &    FSN     &    79.50    &    33.08   &    30.21   &     0.23  &        9.42  &      1.34   \\
         100  &     2.95  &      99.99  &    FSN     &     99.50  &      32.96  &      31.64   &     0.16  &        9.43  &      1.34 \\
         200  &     2.37  &     199.99   &    PISN    &      -    &     110.08   &     106.15   &     0.80  &        9.19  &      1.33 \\
         300  &     2.14  &     299.99   &    DBH     &    299.50  &     169.26   &     164.48   &     0.79  &        9.14  &      1.33 \\
         400  &     2.04  &     399.99   &    DBH     &    399.49  &     227.87   &     221.41   &     0.77  &        9.11  &      1.33 \\
         500  &     1.94  &     499.98   &    DBH     &    499.48  &     286.85   &     277.39   &     0.75  &        9.09  &      1.33 \\
         600  &     1.90  &     599.97   &    DBH     &    599.47  &     345.96   &     333.83   &     0.74  &        9.08  &      1.33 \\
        1000 & 1.80& 999.81 & DBH & 9993.1 &  576.94 &  538.72 &   0.70  &        9.02  &      1.33 \\
        2000 & 1.65  & 1997.26 & DBH & 1996.76 & 1174.06 & 1103.56 & 0.61  &        8.98  &      1.33  \\
       
        \midrule
        
        \multicolumn{10}{c}{$Z = 10^{-6}$} \\
        \midrule 
          14  &     14.09   &    13.99  &    CCSN    &     1.49  &     4.54  &     2.79  &     0.00     &        9.35  &      1.41    \\
          20  &     9.08  &    19.99  &    CCSN    &     1.59  &     7.32  &     5.15  &     0.06  &        9.38  &      1.42  \\
          30  &     6.16  &    29.99  &    FSN     &    29.49   &    12.75   &    10.09   &     0.00     &        9.41  &      1.41  \\
          40  &     4.91  &    39.99  &    FSN     &    39.49   &    18.15   &    15.25    &     0.00     &        9.44  &      1.38  \\
          50  &     4.28  &    49.99  &    FSN     &    49.49   &    22.96   &    21.04   &     0.08  &        9.43  &      1.37  \\
          80  &     3.37  &    79.98   &    FSN     &    79.48    &    28.84   &    26.32   &     0.04  &        9.40  &      1.35  \\
        100  &     3.06  &      99.96  &    PPISN   &     43.10  &      49.92  &      44.12  &     0.66  &        9.36  &      1.33 \\
         200  &     2.46  &     199.93   &    PISN    &      -    &     111.07   &     103.03   &     0.80  &        9.18  &      1.33 \\
         300  &     2.24  &     299.73   &    DBH     &    299.23  &     173.02   &     164.90   &     0.79  &        9.13  &      1.33 \\
         400  &     2.12  &     399.24   &    DBH     &    398.74  &     234.43   &     223.38   &     0.78  &        9.10  &      1.33  \\
         500  &     2.05  &     498.48   &    DBH     &    497.98  &     294.70   &     283.18   &     0.76  &        9.08  &      1.33 \\
         600  &     1.99  &     597.55   &    DBH     &    597.05  &     352.42   &     338.47   &     0.74  &        9.08  &      1.33 \\
        1000  &     1.88  &     989.94   &    DBH     &   989.45 &     593.16   &     567.88   &     0.69  &        9.04  &      1.33 \\
        2000  &     1.76  &    1950.40   &    DBH     &   1981.27  &    1169.90   &    1106.60   &     0.61  &        8.78  &      1.34 \\
        \midrule
        \multicolumn{10}{c}{$Z = 0.0001$} \\
        \midrule
         14  &     14.47 &    13.99  &    CCSN    &     1.49  &     4.91  &     2.84  &     0.00     &        9.35  &      1.43  \\
          20  &     9.29  &    19.97  &    CCSN    &     1.57   &     7.56   &     4.98   &     0.02  &        9.40  &      1.42  \\
          30  &     6.29  &    29.95  &    FSN     &    29.45   &    12.65   &     9.58  &     0.04  &        9.38  &      1.41  \\
          40  &     5.04  &    39.90  &    FSN     &    39.40   &    18.44   &    14.88   &     0.00     &        9.43  &      1.39  \\
          50  &     4.37  &    49.84  &    FSN     &    49.33   &    24.48   &    20.44   &     0.02  &        9.44  &      1.37  \\
          80  &     3.42  &    78.88  &    PPISN   &    38.69   &    42.83   &    37.68   &     0.53  &        9.39  &      1.34  \\
        100  &     3.15  &      97.18  &    FSN     &     96.68  &     32.78  &     32.45  &     0.06  &        9.39  &      1.35 \\
         200  &     2.51  &     196.45   &    PISN    &      -     &    113.17   &    104.09   &     0.80  &        9.18  &      1.33 \\
         300  &     2.29  &     291.76   &    DBH     &    291.26   &    174.97   &    160.07   &     0.79  &        9.12  &      1.33 \\
         400  &     2.17  &     384.07   &    DBH     &    383.57   &    231.80   &    212.58   &     0.78  &        9.10  &      1.33 \\
         500  &     2.09  &     473.87   &    DBH     &    473.38   &    296.34   &    273.56   &     0.76  &        9.08  &      1.33 \\
         600  &     2.01  &     560.44   &    DBH     &    559.94   &    359.77   &    336.05   &     0.74  &        9.06  &      1.33 \\
        1000  &     1.92  &     912.27   &    DBH     &   911.78   &    820.52   &    396.31   &     0.70  &        9.03  &      1.34 \\
        2000  &     1.75  &    1777.3    &    DBH     &   1776.89   &    553.51   &    553.18   &     0.55  &        9.11  &      1.33 \\
        \bottomrule
\end{tabular} 
\end{center}
\label{tab:stellar_properties}
\end{table*}

\begin{table*} 
    
    \caption{Continued} 
    \begin{center}
    \begin{tabular}[c]{c c c c c c c c c c }
    \toprule
    \Mz\ & $t_\mathrm{lifetime}$ &  $M_\mathrm{Pre-SN}$  & Fate & $M_\mathrm{Remnant}$ & $M_\mathrm{He}$ & $M_\mathrm{CO}$ &  $X\mathrm{^c_{O}}$ & $\log T\mathrm{_c}$ & $\langle \Gamma_1 \rangle_\mathrm{Pre-SN}$  \\
        
        [\msun] & [Myr] & [\msun]  & &[\msun] & [\msun] & [\msun]   &   & [K] &  \\
        (1) & (2) & (3) & (4) & (5) & (6) & (7) & (8) & (9) & (10) \\
        \midrule

        \multicolumn{10}{c}{$Z = 0.001$} \\
        \midrule
         14  &     14.65 &    13.94  &    CCSN    &     1.42  &     4.94   &     2.83  &     0.01  &        9.33  &      1.43  \\
          20  &     9.40  &    19.82  &    CCSN    &     1.64  &     7.62  &     4.93  &     0.03  &        9.39  &      1.42  \\
          30  &     6.37  &    29.47  &    FSN     &    28.96   &    12.61   &     9.43  &     0.03  &        9.38  &      1.41   \\
          40  &     5.14  &    38.72  &    FSN     &    38.22   &    18.16   &    14.54   &     0.00  &        9.43  &      1.39  \\
          50  &     4.39  &    46.71   &    FSN     &    46.21   &    24.09   &    19.98   &     0.14  &        9.41  &      1.37   \\
          80  &     3.51  &    69.36  &    PPISN   &    38.78   &    30.59   &    27.65   &     0.04  &        9.39  &      1.35  \\
        100  &     3.19  &     75.08  &    PPISN   &     33.92   &     34.57  &     34.56  &     0.18  &        9.41  &      1.35 \\
         200  &     2.55  &    157.60  &    PISN    &      -      &     89.75  &     85.92  &     0.82  &        9.21  &      1.33  \\
         300  &     2.31  &    213.81  &    DBH     &    213.31    &    174.64   &    157.77   &     0.80  &        9.12  &      1.33 \\
         400  &     2.20  &    289.49  &    DBH     &    288.99    &    222.66   &    206.17   &     0.78  &        9.11  &      1.33 \\
         500  &     2.12  &    363.47  &    DBH     &    362.97    &    274.31   &    255.43   &     0.77  &        9.09  &      1.33 \\
         600  &     2.06  &    439.78  &    DBH     &    439.28    &    323.54   &    303.96   &     0.75  &        9.07  &      1.33 \\
        \midrule
        \multicolumn{10}{c}{$Z = 0.006$} \\
        \midrule
          14  &     14.57  &    13.74  &    CCSN    &     1.37  &     4.78  &     2.69  &     0.01  &        9.36  &      1.43  \\
          20  &     9.31  &    18.08 &    CCSN    &     1.54  &     7.49  &     4.88  &     0.02  &        9.39  &      1.42  \\
          30  &     6.27  &    24.06  &    FSN     &    23.55   &    12.46   &     9.29  &     0.03  &        9.38  &      1.41  \\
          40  &     5.04  &    28.23  &    FSN     &    27.72   &    17.73   &    14.12   &     0.68  &        9.11  &      1.40  \\
          50  &     4.33  &    27.99  &    FSN     &    27.49   &    23.33   &    19.38   &     0.44  &        9.37  &      1.37  \\
          80  &     3.43  &    41.11  &    PPISN   &    38.70   &    40.28   &    35.82   &     0.85  &        9.26  &      1.35  \\
        100  &     3.17  &     52.08  &    PPISN   &     43.84  &     51.56  &     45.51  &     0.86  &        9.27  &      1.34  \\
         200  &     2.54  &    104.99   &    PISN    &      -    &    103.94   &     93.45  &     0.79  &        9.19  &      1.33 \\
         300  &     2.32  &    158.24   &    DBH     &    157.74  &    156.66   &    142.25   &     0.79  &        9.14  &      1.33 \\
         400  &     2.20  &    210.69   &    DBH     &    210.19  &    208.59   &    190.50   &     0.78  &        9.10  &      1.33 \\
         500  &     2.12  &    263.66   &    DBH     &    263.16  &    261.02   &    238.98   &     0.77  &        9.09  &      1.33 \\
         600  &     2.07  &    314.80   &    DBH     &    314.30  &    311.65   &    285.89   &     0.75  &        9.07  &      1.33 \\
        
        \midrule
        \multicolumn{10}{c}{$Z = 0.01$} \\
        \midrule
        14  &     14.30  &    13.53  &    CCSN    &     1.42  &     4.76  &     2.71  &     0.01  &        9.36  &      1.43  \\
          20  &     9.15  &    18.52  &    CCSN    &     1.54  &     7.25  &     4.83  &     0.04  &        9.39  &      1.42   \\
          30  &     6.14  &    23.13  &    FSN     &    22.63   &    12.42   &     9.27  &     0.03  &        9.38  &      1.41   \\
          40  &     5.04  &    20.98  &    FSN     &    20.77   &    17.62   &    14.14   &     0.00  &        9.43  &      1.39  \\
          50  &     4.30  &    19.96  &    FSN     &    19.46   &    19.76   &    17.08   &     0.27  &        9.39  &      1.39  \\
          80  &     3.40  &    33.29  &    FSN     &    32.79   &    32.62   &    28.73   &     0.71  &        9.37  &      1.35  \\
         100  &     3.15  &     38.54  &    PPISN   &     37.53  &     38.15  &     33.69  &     0.61  &        9.38  &      1.34 \\
         200  &     2.55  &     73.60  &    PISN    &      -     &     72.87  &     64.99  &     0.85  &        9.24  &      1.33 \\
         300  &     2.33  &    108.34   &    PISN    &      -     &    107.26   &     96.79  &     0.79  &        9.18  &      1.33 \\
         400  &     2.22  &    141.49   &    DBH     &    140.99   &    140.07   &    127.20   &     0.79  &        9.15  &      1.33 \\
         500  &     2.14  &    172.54   &    DBH     &    172.05   &    170.82   &    155.66   &     0.78  &        9.12  &      1.33  \\
         600  &     2.10  &    186.83   &    DBH     &    186.33   &    184.96   &    168.80   &     0.78  &        9.12  &      1.33  \\
        \midrule
        \multicolumn{10}{c}{$Z = 0.02$} \\
        \midrule
         14  &     13.86  &    13.10   &    CCSN    &     1.49  &     4.60  &     2.69  &     0.01  &        9.36  &      1.43  \\
          20  &     8.75  &    17.74   &    CCSN    &     1.55  &     7.10  &     4.85  &     0.68  &        9.20  &      1.43  \\
          30  &     5.91  &    13.21   &    FSN     &    12.70   &    12.81   &     9.32   &     0.05  &        9.38  &      1.41  \\
          40 & 4.90 & 8.93  & FSN & 8.43 & 8.93 & 7.23 & 0.00 & 9.42 & 1.40\\
          50  &     4.21  &    11.84   &    FSN     &    11.33   &    11.59   &     9.64  &     0.17  &        9.37  &      1.39  \\
          80  &     3.39  &    15.86   &    FSN     &    15.36   &    15.61   &    13.09   &     0.11  &        9.40  &      1.39  \\
         100  &     3.18  &    17.38  &    FSN     &    16.88  &    17.21  &    14.74  &     0.04  &        9.41  &      1.38  \\
         200  &     2.65  &    25.05  &    FSN     &    24.55  &    24.80  &    21.02  &     0.50  &        9.38  &      1.36  \\
         300  &     2.44  &    33.19   &    FSN     &    32.69   &    32.86  &    28.63  &     0.61  &        9.38  &      1.35  \\
         400  &     2.35  &    34.97  &    PPISN   &    34.32  &    34.62  &    30.35  &     0.67  &        9.38  &      1.34  \\
         500  &     2.33  &    34.79  &    PPISN   &    34.14  &    34.45  &    30.14  &     0.68  &        9.37  &      1.34  \\
         600  &     2.32  &    34.77  &    PPISN   &    34.12  &    34.42   &    30.15  &     0.63  &        9.38  &      1.34  \\
        \bottomrule

\end{tabular} 
\end{center}
\footnotesize{Column 1: Initial mass. Column 2: Stars lifetime. Column 3: Final mass. Column 4: Final fate of the star. Column 5: Remnant mass. Columns 6 and 7: He and CO core mass at the end of computation. Column 8: oxygen central mass fraction at the end of the computation. Column 9: Central temperature of the last model. Column 10: $\langle \Gamma_1 \rangle_\mathrm{Pre-SN}$ at the end of the computation}. 
\label{tab:stellar_properties2}
\end{table*}

\end{document}